\documentclass[5p]{elsarticle}

\usepackage{color}
\usepackage{xcolor}
\usepackage{array}
\newcolumntype{x}[1]{>{\centering\arraybackslash\hspace{0pt}}p{#1}}
\usepackage{colortbl}
\usepackage{url}
\usepackage{verbatim}
\usepackage{multirow}
\usepackage{xspace}
\usepackage{graphicx}
\usepackage{amsmath}
\usepackage{amssymb}
\usepackage{booktabs}
\usepackage{times}
\usepackage{enumitem}
\usepackage{xcolor}
\usepackage[colorlinks=true,linkcolor=blue,citecolor=blue]{hyperref}%

\newcommand{\hide}[1]{}

\newcommand{\eg}{\emph{e.g.}\xspace}
\newcommand{\ie}{\emph{i.e.}\xspace}
\newcommand{\etal}{\emph{et al.}\xspace}

\let\oldnl\nl
\newcommand{\nonl}{\renewcommand{\nl}{\let\nl\oldnl}}
\usepackage[normalem]{ulem}
\newcolumntype{L}[1]{>{\raggedright\let\newline\\\arraybackslash\hspace{0pt}}m{#1}}

\usepackage{todonotes}

\newcommand{\rev}[1]{{\color{black}#1}}

\usepackage{lineno}
\usepackage{pdftexcmds}
\usepackage[shortcuts]{extdash}
\modulolinenumbers[5]

\begin{document}

\begin{frontmatter}

\title{Machine Learning for Integrating Data in Biology and Medicine:\\ Principles, Practice, and Opportunities}

\author[stanford,corr]{Marinka Zitnik}
\ead{marinka@cs.stanford.edu}
\author[toronto-biophysics,toronto-cancer]{Francis Nguyen} 
\author[bo]{Bo Wang}
\author[stanford,biohub,corr]{Jure Leskovec}
\ead{jure@cs.stanford.edu}
\author[toronto-sickchildren,toronto-cs,toronto-vector,corr]{Anna Goldenberg}
\ead{anna.goldenberg@utoronto.ca}
\author[toronto-biophysics,toronto-cancer,toronto-cs,toronto-vector,corr]{Michael M.~Hoffman} 
\ead{michael.hoffman@utoronto.ca}

\address[stanford]{Department of Computer Science, Stanford University, Stanford, CA, USA}
\address[toronto-biophysics]{Department of Medical Biophysics, University of Toronto, Toronto, ON, Canada}
\address[toronto-cancer]{Princess Margaret Cancer Centre, Toronto, ON, Canada}
\address[bo]{Hikvision Research Institute, Santa Clara, CA, USA}
\address[biohub]{Chan Zuckerberg Biohub, San Francisco, CA, USA}
\address[toronto-sickchildren]{Genetics \& Genome Biology, SickKids Research Institute, Toronto, ON, Canada}
\address[toronto-cs]{Department of Computer Science, University of Toronto, Toronto, ON, Canada}
\address[toronto-vector]{Vector Institute, Toronto, ON, Canada\\[2mm]}
\address[corr]{{\em Corresponding authors}}

\begin{abstract}
New technologies have enabled the investigation of biology and human health at an unprecedented scale and in multiple dimensions.
These dimensions include \rev{a myriad of} properties describing genome, epigenome, transcriptome, microbiome, phenotype, and lifestyle.
No single data type, however, can capture the complexity of all the factors relevant to understanding a phenomenon such as a disease.
Integrative methods that combine data from multiple technologies have thus emerged as critical statistical and computational approaches.
The key challenge in developing such approaches is the identification of effective models to provide a comprehensive and relevant systems view.
An ideal method can answer a biological or medical question, identifying important features and predicting outcomes, by harnessing heterogeneous data across several dimensions of biological variation.
In this Review, we describe the principles of data integration and discuss current methods and available implementations.
We provide examples of successful data integration in biology and medicine.
Finally, we discuss current challenges in biomedical integrative methods and our perspective on the future development of the field.
\end{abstract}


\end{frontmatter}

\section{Introduction}\label{sec:intro}

\begin{figure*}[t]
\centering
\includegraphics[width=\linewidth]{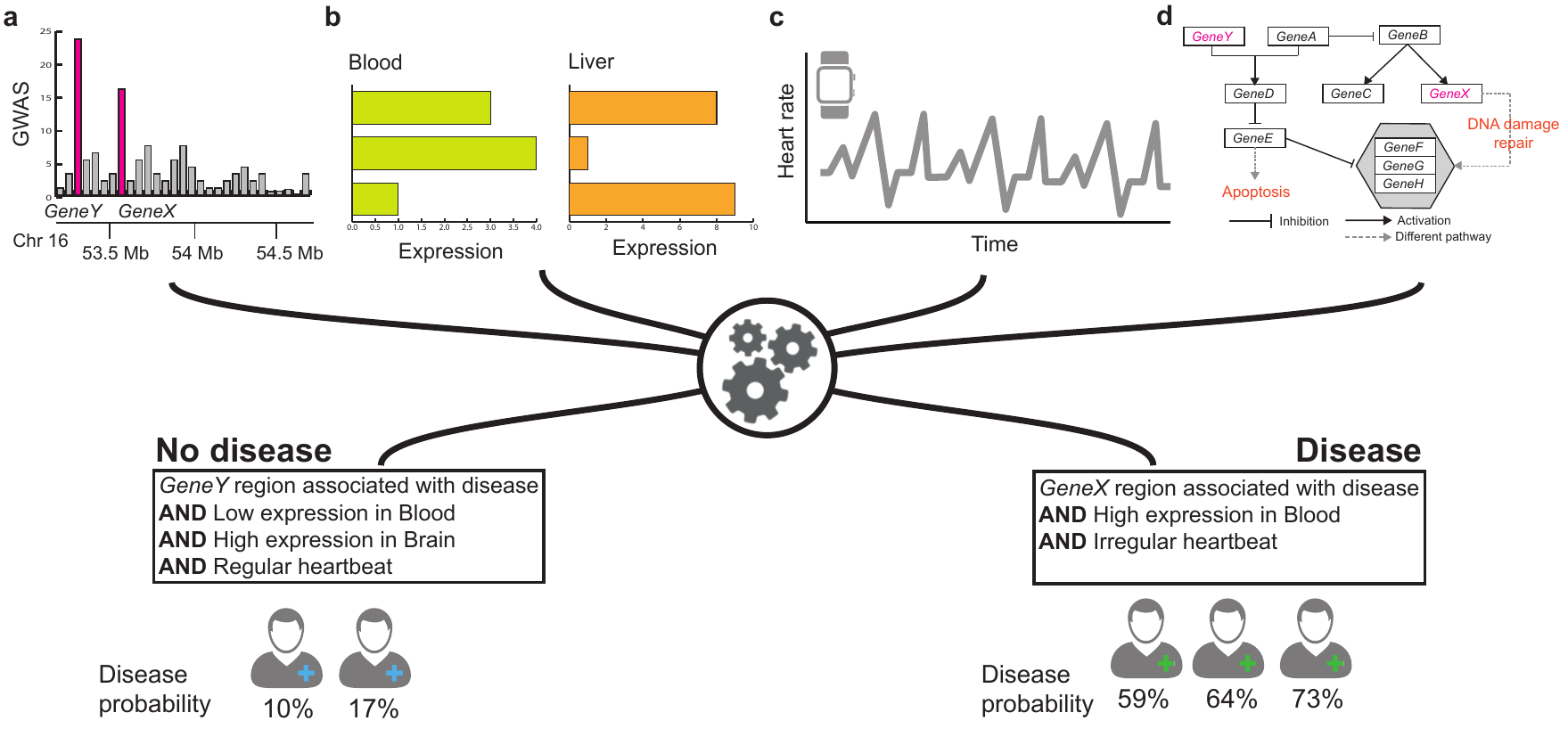}
\caption{\textbf{The importance of data integration in biomedicine.} Considering variation in only a single data type can miss many important patterns that can only be observed by  considering multiple levels of biomedical data. Shown is a hypothetical example using disease diagnostics as a point of interest. When a new patient arrives to the clinic, (\textbf{a}) domain experts sequence the patient's genome and compare it with a database to identify mutations and disease-causing genes, (\textbf{b}) perform laboratory tests using tissue samples, and (\textbf{c}) process information about the patient's behavior and lifestyle. (\textbf{d}) The patient's genomic, transcriptomic, and lifestyle information is combined with curated databases of biomedical knowledge (\eg, disease and metabolic pathways). Finally, a machine learning algorithm predicts probability that the patient will develop a particular disease in near future. To make accurate prediction, the machine learning model needs to use many different types of data. This example illustrates that accurate prediction can only be made by analyzing multiple types of patient's data.}
\label{fig:why-integrate}
\end{figure*}

Understanding complex biological systems has been an ongoing quest for many researchers.
The rapidly decreasing costs of high-throughput sequencing, development of massively parallel technologies, and new sensor technologies have enabled \rev{generation of data that describe biological systems on multiple dimensions.}
\rev{These} dimensions include DNA sequence~\cite{Encode2012integrated}, epigenomic stat\rev{es}~\cite{Kundaje2015IntegrativeEpigenomes}, single-cell \rev{gene} expression activity~\cite{Quake2018single}, proteomics~\cite{Wilhelm2014mass}, functional and phenotypic measurements~\cite{Costanzo2016global}, and ecological and lifestyle properties~\cite{Li2017digital}.
These technological advances in data generation have driven the field of bioinformatics for the past decade, producing ever increasing amounts of data \rev{of different types} as researchers develop \rev{data analysis} tools.
Many of these data types have associated analytical methods designed to examine one data type specifically.
Using these methods, \rev{studies} have assembled some of the puzzle of biological architecture.
Usually, however, the factors necessary to understand a phenomenon such as a disease, cannot be captured by a single data type (\autoref{fig:why-integrate}).
Much of the complexity in biology and medicine thus remains unexplained.
If \rev{the field relies} strictly on single-data-type studies, it never will be explained.

Ideally, \rev{one} can combine different types of data to create a holistic picture of the cell, human health, and disease.
Researchers have developed multiple approaches to do this, and therefore address the challenges brought forward by large and heterogeneous biomedical data.
For example, one can identify DNA sequence variation through association studies in family- and population-based data, and then integrate it with molecular pathway information to predict the risk of developing a particular disease~\cite{Chatterjee2013projecting}.
\rev{Data integration can have numerous meanings, however, it is used here to mean} the process by which different types of biomedical data in their broadest sense are combined as predictor variables to allow for more thorough and comprehensive modeling of biomedically relevant outcomes.
As reviewed previously (\eg, \cite{Ritchie2015methods,Karczewski2018integrative,Teschendorff2018review}), a data integration approach can achieve a more thorough and informative analysis of biomedical data than an approach that uses only a single data type.
Combining multiple data types can compensate for missing or unreliable information in any single data type, and multiple sources of evidence pointing to the same outcome are less likely to lead to false positives.
\rev{A complete model of a system like the human body is only likely to be discovered if information from different dimensions is considered,} from the genome and transcriptome to organismal environment.

In this Review, we describe the principles of data integration, and provide a taxonomy of machine learning methods pre\-sen\-tly in use to integrate biomedical data.
We discuss current methods, implementations of these methods, and their successful applications in biology and medicine.
Furthermore, we discuss challenges in optimally combining and interpreting data from multiple sources and the advantages of integrating multiple data types.
For example, one technology may address shortcomings of another to provide a more precise insight into human disease.
In addition, we provide our perspective on how integrative data analysis might develop in the future.

\section{Challenges in data integration for biology and medicine}

When one develops machine learning approaches to integrate biomedical data, several challenges arise.
Biological and medical datasets have inherent complexity beyond their large sizes.
Biomedical datasets are also high-dimensional, incomplete, biased, heterogeneous, dynamic, and noisy.
We briefly describe these challenges below.

Biomedical data is often high-dimensional but sparse.
This contrasts with large datasets in other domains, such as social networks, computer vision, and natural language, that typically contain a large number of high-quality examples.
A typical geno\-me-wide association study (GWAS)~\cite{Hu2016gwas} genotypes hundreds of thousands of single-nucleotide polymorphisms for every individual.
However, these data can often be collected for only a relatively small number of individuals with a particular phenotype.
Furthermore, the sparse nature of these data, \ie, each polymorphism is only present in a small number of all individuals, presents an additional challenge for downstream analytic applications.
It remains a major challenge to convert these data into biologically and clinically meaningful insights.
Without integrating other types of data, such as pathway or molecular network information~\cite{Linghu2009genome,Hofree2013network,Lundby2014annotation}, GWAS data alone can struggle to identify meaningful patterns associated with the phenotype of interest.

Another important challenge arises from the often incomplete and biased nature of biomedical data.
This challenge \rev{ari\-ses} from  limitations of measurement technology~\cite{Zitnik2015impute}, natural and physical constraints~\rev{\cite{Hu2016gwas,Hyde2016identification}}, and investigative biases~\cite{Menche2015}.
For example, \rev{information on what chemical compounds bind to what genes is available for only several thousands of genes, even when considering information from across organisms}~\cite{Campillos2008}.
Furthermore, the number of associated compounds for each gene is highly uneven~\cite{Zong2017deep}, with many uncharacterized genes playing important roles in drug action~\cite{Hodos2016}.
Additionally, biome\-dical data are hierarchically organized and span molecules, pathways, cells, tissues, organs, patients, and populations~\cite{Carvunis2014siri,Greene2015understanding,Zitnik2017ohmnet} and also cover a wide spectrum of time\-sca\-les and species.
Clearly, full understanding of biology requires multiscale modeling, from describing atomic details of mo\-le\-cules to the emergent properties of organismal populations.
Furthermore, when biomedical outcomes change over time, machine learning methods integrating the outcomes need to account for these dynamics.
For example, cancer cells, bacteria, and viruses evolve rapidly to gain drug \rev{resistance~\cite{Bicker2017},} and ignoring the dynamics of drug response can lead to poor performance in predicting drug efficacy and toxicity.

A fundamental challenge in biomedical data science lies in discovering new knowledge outside of the existing domain of know\-ledge, \eg, extrapolating a drug response from an animal model to that in a human patient.
Existing approaches typically assume that the dataset on which the algorithm is trained is representative of all the data to which the algorithm can be applied.
However, it is challenging to build a model to predict, \eg, efficacy of an anticancer drug in a given patient, as a new patient might be unique and might fall outside of the hypothesis space of the trained model.
\rev{A}s biomedical datasets are incomplete and reflect scientific knowledge discovered so far, the models can be trained on only these partially complete datasets and thus can perform poorly when new data become available.
For these reasons, it is especially challenging to deploy machine learning systems to support decision making in risk-sensitive discovery and clinical practice~\cite{Mullainathan2017does}, \eg, the system might make conflicting predictions about \rev{the} utility of a particular anticancer drug for a given patient depending on the type of input data used for prediction.

In summary, due to the complex and interconnected nature of biomedical systems, any single model trained on any single dataset can touch only a small part of the entire biomedical knowledge.
It is thus critical to integrate diverse sources of information to gain a comprehensive understanding of biology and medicine.

\section{Conceptual organization of methods for data integration}

\begin{figure}[t]
\centering
\includegraphics[width=\linewidth]{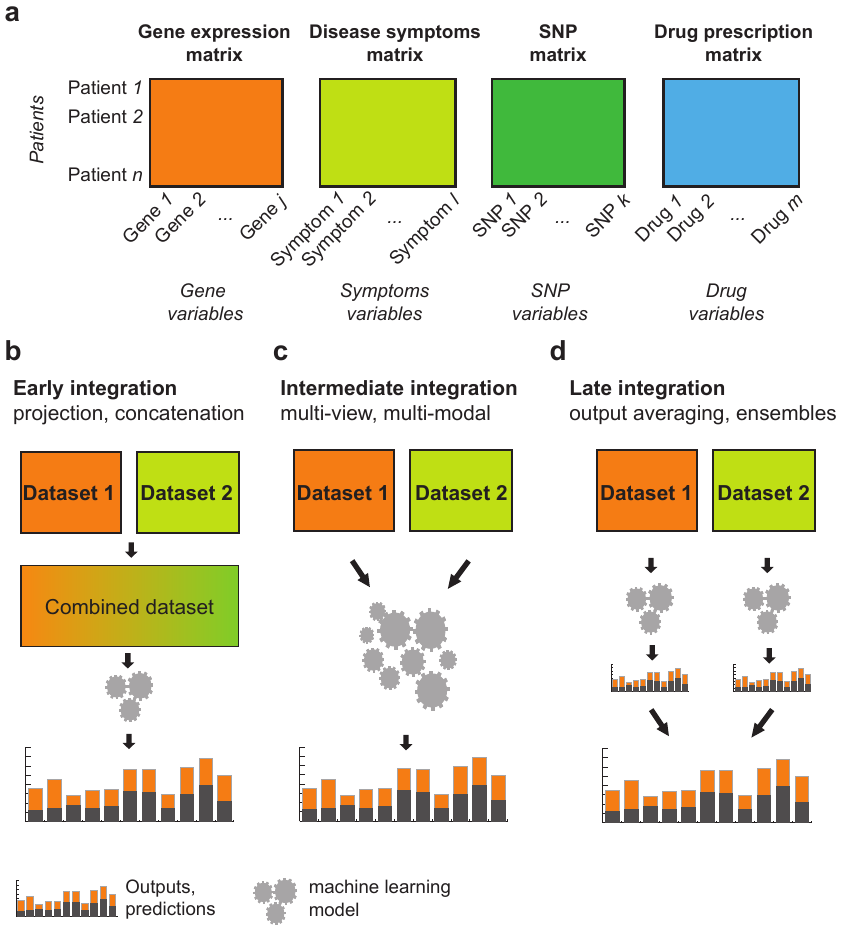}
\caption{\textbf{Categorization of approaches for data integration.} (\textbf{a}) Examples of multi-omics data about patients. (\textbf{b-d}) Data integration approaches can be divided into three categories. (\textbf{b}) {\em Early integration approaches} involve combining datasets from different data types at the raw or processed level before analysis and prediction. (\textbf{c}) {\em Intermediate integration approaches} transform or map the underlying datasets at the same time as they estimate model parameters. (\textbf{d})~{\em Late integration approaches} perform analysis on each dataset independently, which is followed by integration of the resulting models to generate predictions, \eg, prognosis for a particular patient. SNP, single-nucleotide polymorphism.}
\label{fig:integration-strategy}
\end{figure}

We broadly categorize data integration methods into two types of approaches.
We refer to approaches that combine models and datasets across \rev{ spatial and temporal scales as {\em vertical data integration}, which depends on integration of cellular, cell type, tissue, organism, and population models at several temporal scales~\cite{Zitnik2017ohmnet,Pilosof2017multilayer,Zitnik2016jumping}. In contrast, {\em horizontal data integration} focuses on combining datasets and models at one particular level~\cite{Bujold2016ThePortal,Libbrecht2015JointExpression}, for example, at the microbiome~\cite{Smits2017seasonal} or at the epigenome level~\cite{Kundaje2015IntegrativeEpigenomes}.}

More technically, the methods implement one of the following three distinct approaches to data integration depending on the \rev{analysis} stage at which integration takes place~\rev{\cite{Ritchie2015methods,Pavlidis2002learning,Maragos2008cross,Zitnik2015data}} (\autoref{fig:integration-strategy}).
{\em Early integration} (\autoref{fig:integration-strategy}\rev{b}) begins by transforming all datasets into a single, feature-based table or a graph-based \rev{representation, which is then used as input} to a machine learning method.
\rev{Theoretically, this approach is powerful because the machine learning method} can consider any type of dependence between the features as long as individual datasets are not collapsed \rev{before analysis}.
Early integration approaches often rely on methods for automatic feature learning, such as dimensionality reduction~\cite{Zitnik2012nimfa} and representation learning~\rev{\cite{Vincent2010stacked,Sarajlic2016graphlet,Hamilton2017review}}, to project raw high-dimensional datasets into a low-dimensional vector space and then combine these low-dimensional representations through concatenation or other simple aggregation techniques.

\rev{I}n {\em late integration} (\autoref{fig:integration-strategy}\rev{d}), a first-level model is built for each dataset or data type independently.
These first-level models are then combined by training a second-level model that uses predictions of the first-level models as features or via a meta-predictor~\cite{Yang2010review} that takes a majority vote or combines prediction weights of the first-level models~\cite{Wu2010prediction,Iam2010lce}.

\rev{In {\em intermediate integration} (\autoref{fig:integration-strategy}\rev{c}), a model, such as multiple kernel} learning~\cite{Brayet2014towards,Mariette2017unsupervised}, collective matrix factorization~\cite{Zitnik2014survival,Wang2014similarity,Zitnik2015data} or deep neural network~\cite{Tan2017unsupervised,Zitnik2018polypharmacy}, learns a joint representation of many datasets.
Intermediate integration relies on algorithms \rev{that explicitly address the multiplicity} of datasets and fuse them through inference of a joint model.
Importantly, \rev{an intermediate data integration method} does not combine input data nor does it develop a separate model for each dataset.
Instead, it aims to preserve the structure of data and only merge them during the \rev{analysis} stage.
The intermediate integration approach can lead to superior performance, however it often requires development of a new algorithm and cannot be used with off-the-shelf software tools.

\begin{figure}[t]
\centering
\includegraphics[width=\linewidth]{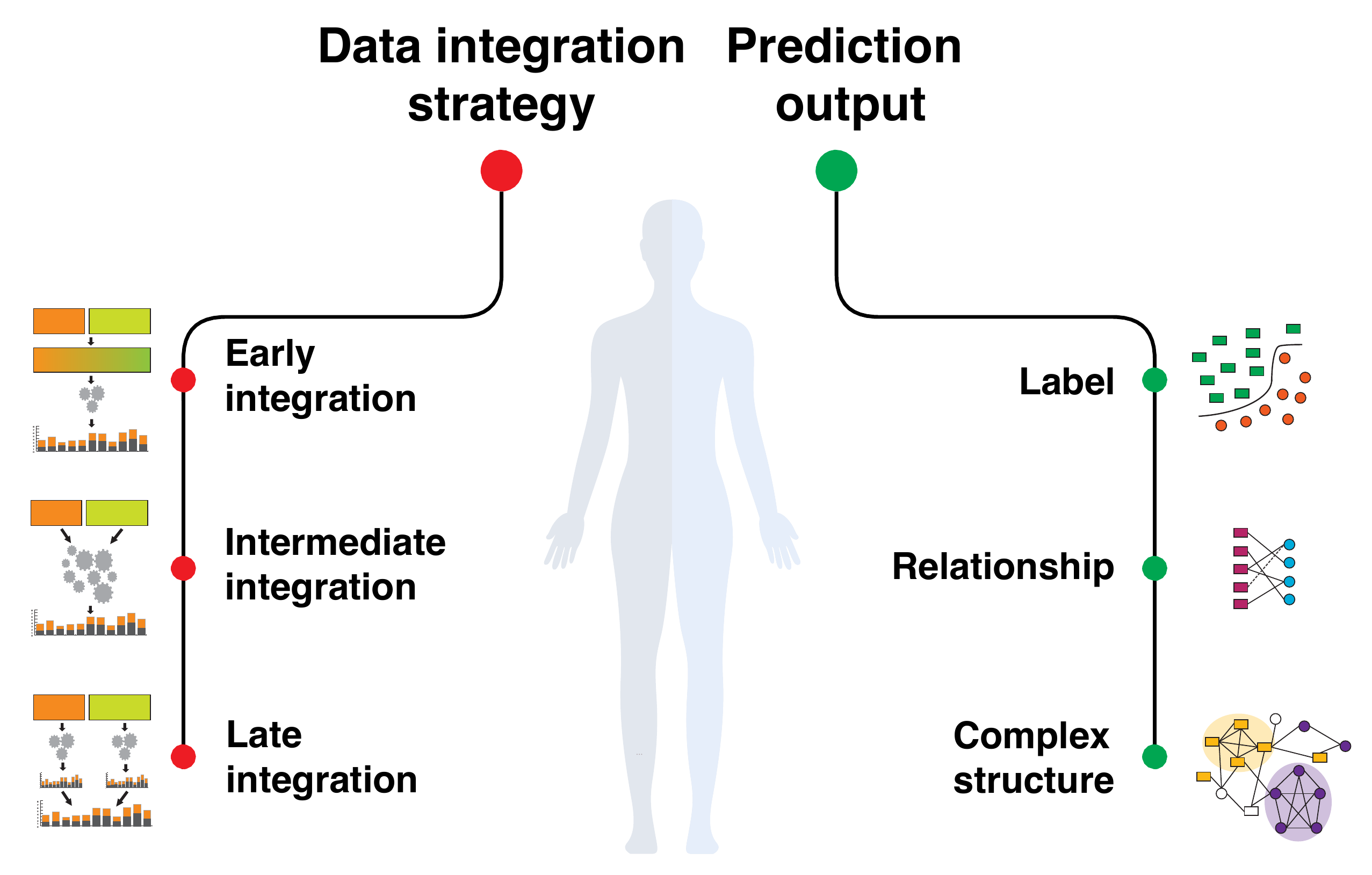}
\caption{\textbf{Data integration.} Data integration approaches combine multiples sources of information in a statistically meaningful way to provide a comprehensive analysis of \rev{biomedical data}. Broadly, existing approaches \rev{use three distinct strategies} (\ie, early, intermediate, and late integration; see also \autoref{fig:integration-strategy}) and produce three types of prediction outputs (\ie, a label representing probability of an entity belonging to a given class; a relationship representing probability of an association between two entities; and a complex structure, such as an inferred network or a partitioning of entities into groups).}
\label{fig:integration-strategy-prediction-type}
\end{figure}

Finally, methods for data integration can generate diverse types of prediction outputs similar to methods that \rev{analyze a single dataset} (\autoref{fig:integration-strategy-prediction-type}).
One area of a particular interest is the prediction \rev{of  quantitative or categorical properties ({\em labels}, \eg, gene functions) for biomedical entities (\eg, genes)}.
For example, many studies \rev{integrate a large number of networks}, including protein-protein and genetic interaction networks, which are now available for several organisms, to predict genes that cause a particular phenotype or have a particular function~\cite{Mostafavi2008genemania,Carreras2017comprehensive} (Section~\ref{sec:protein-function-prediction}).
Beyond predicting labels of individual entities, many studies aim to predict {\em relationships}, \ie, molecular interactions, functional associations, or causal relationships between biomedical entities.
For example, a multiple kernel learning approach can combine kernels derived from diverse data, such as \rev{a} drug's structural similarity, \rev{a} drug's phenotypic similarity, and target similarity, to predict new {\em relationships} between a drug and proteins that the drug might target~\cite{Gonen2012predicting}, \ie, drug-target interactions (Section~\ref{sec:drug-target-interaction-prediction}).
\rev{Finally, data integration methods exist to} identify {\em complex structures}, such as gene modules or clusters detected in \rev{a} combined gene interaction network~\cite{Cowen2017network} (Section~\ref{sec:protein-protein-interaction-prediction}), and to generate structured outputs, such as gene regulatory networks inferred from \rev{mixed data} distributions~\cite{Zitnik2015netinf}.


\section{Focus of this \rev{Review}}

This Review is intended for computational researchers \rev{interested in} recent developments and applications of machine learning to biology and medicine and its potential for advancing biomedicine given the vast amounts of heterogeneous data being generated today.
In the Review, we focus on statistical approaches and machine learning methods for data integration.
We describe the principles of integrative approaches and provide an overview of some of the \rev{methods used to address various biomedical questions}, the tools available to implement these analyses, and the various strengths and weaknesses of integrative approaches.
Additionally, we highlight outstanding challenges and opportunities that are ripe for exploration using \rev{new machine learning methods and provide} our perspective on how integrative approaches might develop in the future.

Several reviews cover related data integration topics from different perspectives, or with a special focus on a particular \rev{biomedical question}.
For example, Rider \etal~\cite{Rider2013review} focus on methods for network inference with a special focus on probabilistic methods.
Bebek \etal~\cite{Bebek2012review} and Cowen \etal~\cite{Cowen2017network} focus on methods for construction and statistical analysis of biological networks from multiple biological datasets, as well as on visualization tools.
Related \rev{reviews \cite{Ritchie2015methods,Kristensen2014review,Gligorijevic2016review,Malod2017precision}} survey recent advances in high-throughput technologies and data integration-based methods for translational medicine and list the tools that are available to domain scientists.
Karczewski \etal~\cite{Karczewski2018integrative} describe applications of data integration that combine diverse types of data to understand, diagnose and inform treatment of diseases.
They discuss technical challenges \rev{of implementing integrative approaches} in clinics and for personalized medicine.
Teschendorff \etal~\cite{Teschendorff2018review} \rev{survey} algorithms for drawing inferences from biological sequence data with a focus on statistical analysis of genome sequencing data.

In this Review, we survey advances in data integration at multiple \rev{biomedical levels.}
We organize our presentation according to the flow of genetic information from the genome level to the transcriptome level and, ultimately, to the phenome level.
Heterogeneous data exist within and between these levels. We start at the DNA sequence level, describing methylation patterns and other epigenetic markers (Section~\ref{sec:epigenomic} and Section~\ref{sec:noncoding}), proceed at the single-cell level of gene expression (Section~\ref{sec:singlecell}), protein variation and cellular phenotypes (Section~\ref{sec:function}), and reach the patient population levels (Section~\ref{sec:drug} and Section~\ref{sec:patient}).
Finally, we discuss the potential for combining \rev{diverse data} and the central role of integrative approaches in human health and disease (Section~\ref{sec:discussion}).

\section{Epigenomic variation and gene regulation}\label{sec:epigenomic}

\begin{table*}[t]
  \centering
  \begin{tabular*}{\linewidth}{r@{\extracolsep{\fill}}p{\linewidth*5/7}}
    \toprule
    \textbf{Term} & \textbf{Description} \\
    \midrule
    assay & Laboratory experiment used to measure some physical or chemical aspect of a sample. \\
    chromatin & DNA, its structure for packaging, and the attached biomolecules. \\
    chromatin accessibility & Measurement of the openness of chromatin. \\
    chromatin state & Label summarizing multiple properties of a region of chromatin, which often include histone modifications, chromatin accessibility, and transcription factor binding. \\
    conservation & Measurement of how little a particular sequence changes throughout evolution. \\
    deleterious & Hindering an organism's survival. \\
    DNA methylation & Chemical modification to DNA that alters protein binding affinity without changing sequence. \\
    dual futility conjecture & Conjecture that many transcription factor binding sites cannot be predicted from sequence alone. \\
    epigenome & The collection of site-specific chemical and physical properties of the genome other than its sequence. \\
    enhancer & Genomic region that influences transcription of a gene distant along the one-dimensional chromosome. \\
    futility conjecture & Conjecture that many transcription factor binding sites predicted from sequence alone will have no functional role. \\
    histone & Class of protein that packages DNA into nucleosomes. \\
    histone modifications & Chemical alterations to histones that can alter gene expression. \\
    label & Identifier for a pattern or cluster describing multiple regions of the genome, such as a chromatin state. \\
    motif & Short, recurring sequences recognized by proteins such as transcription factors.
            Often defined probabilistically as a position weight matrix. \\
    noncoding & Occurring outside the protein-coding sequence of any gene. \\
    nucleosome & Eight histones and the DNA wrapped around them. \\
    open chromatin & Region of chromatin not packaged into a nucleosome. Available for binding by other proteins. \\
    position weight matrix & Probabilistic model that scores how well a motif describes a sequence.
    The matrix has a column for each position in the sequence and a row for each symbol in the sequence's alphabet. \\
    regulatory region & Region of DNA with a known effect on gene expression. \\
    segmentation & Partition of the genome with a label assigned to every segment. \\
    topologically associated domain & Region of the genome enriched for three-dimensional interactions within. \\
    transcription factor & Class of protein that binds to chromatin and regulates gene expression. \\
    \bottomrule
  \end{tabular*}
  \caption{\textbf{Epigenomics glossary.}
    Terms referenced in this section.}
  \label{tab:epig-glossary-terms}
\end{table*}

Individual cells within a multicellular organism usually have nearly identical DNA sequences, but still develop distinct cellular identities.
These cellular identities manifest as diverse physical forms and behaviors, but ultimately represent differing programs of gene expression.
The different gene expression programs also materialize in site-specific physical and chemical changes to the DNA and the thousands of biomolecules that interact with it.
These include chemical modification of DNA bases~\cite{Klose2006GenomicMediators,Severin2013EffectsSeparation,Spruijt2014DNATricks}, and of the \emph{histone} proteins that package DNA~\cite{Rothbart2014InterpretingModifications,Stirzaker2014MiningChallenges} into \emph{nucleosome} structures. 
The collection of DNA, its packaging, and associated biomolecules is known as \emph{chromatin}.
Biologists often refer to the state of physical and chemical chromatin changes as a cell's \emph{epi\-geno\-me}~\cite{Lappalainen2017AssociatingPhenotypes} (\autoref{tab:epig-glossary-terms}), and measure its properties base-by-base along the genome.

\begin{table*}[t]
  \centering
  \begin{tabular*}{\linewidth}{r >{\raggedright}p{0.2\linewidth} @{\extracolsep{\fill}}p{0.5\linewidth} l}
    \toprule
    \textbf{Assay} & \textbf{Property measured} & \textbf{Method} & \textbf{References} \\
    \midrule
    ATAC-seq & chromatin accessibility & Uses the Tn5 transposase to insert a short sequence at open chromatin, followed by sequencing. & \cite{Buenrostro2013TranspositionPosition} \\
    BS-seq & DNA methylation & Converts unmethylated cytosines into uracil, followed by sequencing. & \cite{Cokus2008ShotgunPatterning} \\
      CETCh-seq & associated protein & ChIP-seq against a special protein region added by clustered regularly interspaced short palindromic repeats~(CRISPR) genome editing. & \cite{Arnold2013ModelingTargeting, Savic2015CETCh-seq:Proteins} \\
    ChIA-PET & long-range interactions & Ligates DNA regions close in three dimensions together with a known linker sequence, followed by sequencing & \cite{Fullwood2009AnInteractome} \\
    ChIP-exo & associated protein & Like ChIP-seq, with a step that cuts DNA fragments closer to a bound protein. & \cite{Rhee2012ChIP-exoAccuracy} \\
    ChIP-nexus & associated protein & Like ChIP-exo, with an additional self-circularization step that increases library generation efficiency. & \cite{He2015ChIP-nexusFootprints} \\
    ChIP-seq & associated protein & Pulls down regions associated with a protein using an antibody against that protein, followed by sequencing. & \cite{Johnson2007Genome-wideInteractions,Robertson2007Genome-wideSequencing,Barski2007High-ResolutionGenome,Mikkelsen2007Genome-wideCells} \\
    CUT\&RUN & associated protein & Like ChIP-seq, with antibodies that diffuse into the cell to avoid breaking the cell apart prematurely. & \cite{Skene2017AnSites} \\
    DNase-seq & chromatin accessibility & Uses a deoxyribonuclease~(DNase) protein to cut DNA at open chromatin, followed by sequencing. & \cite{Song2010DNase-seq:Cells} \\
    Hi-C & long-range interactions & Ligates DNA regions close in three dimensions together, followed by sequencing. & \cite{Lieberman-Aiden2009ComprehensiveGenome,deWit2012AOrganization} \\
    Hi-ChIP & long-range interactions and associated protein & Combines Hi-C and ChIP-seq. Ligates DNA regions inside of the nucleus with biotin, and applies ChIP-seq on ligated reads. & \cite{Mumbach2016HiChIP:Architecture} \\
    \bottomrule
  \end{tabular*}
  \caption{\textbf{Epigenomic assays.} Glossary of assays referenced in this section.}
  \label{tab:epig-glossary-assays}
\end{table*}

Researchers use investigative experiments known as \emph{assays} to determine epigenomic properties of each region in the genome (\autoref{tab:epig-glossary-assays}).
For example, the histones \rev{that DNA wraps} around can undergo various chemical changes known as \emph{histone modifications}~\cite{Rothbart2014InterpretingModifications}.
The chromatin immunoprecipitation-sequencing (ChIP-seq)~\cite{Johnson2007Genome-wideInteractions,Robertson2007Genome-wideSequencing,Barski2007High-ResolutionGenome,Mikkelsen2007Genome-wideCells} assay can map histone modifications, one at a time.
As another example, nucleosomes often consistently locate at particular DNA regions in particular cell types.
Nucle\-osome-free regions or \emph{open chromatin} play a critical role in the control of gene regulation.
A variety of techniques map nucleosomes and open chromatin, which include deoxyribonuclease-sequencing (DNase-seq)~\cite{Song2010DNase-seq:Cells} and assay for transposase-accessible chromatin (ATAC-seq)~\cite{Buenrostro2013TranspositionPosition}.

Epigenomic sequencing assays usually break genomic DNA into fragments around 200 \rev{base pairs (bp)} in length. 
This fragmentation enriches for chromatin with some epigenomic property of interest, such as a particular histone modification.
These assays end by sequencing the pool of fragments enriched for the sought-after property.
In other kinds of \rev{genomic} sequencing experiments we might find the genetic variation in produced sequencing reads interesting.
Instead, in an epigenomics sequencing assay, we are usually interested primarily in where these reads map in a reference genome---and how often.
For each position in the genome, we can count the number of reads mapped to that position and treat that as a signal of the strength or frequency of the epigenomic property under analysis.
Thus, we can treat the result of the experiment as a numerical vector across the genome.
Usually we include other normalization steps to account for differences in experimental parameters, such as dividing by the total number of mapped reads.
This transforms the initial integer counts into a real-valued vector.
For the human genome at full resolution, this vector would have 3 billion components.

Since epigenomic data might bear only an indirect connection to biological phenomena of interest, machine learning appeals as an aid for interpretation~\cite{Holder2017MachineApplications}.
Researchers have devised numerous ways to draw conclusions about the control of gene expression and its effect \rev{on} phenotype from epigenomic data~\cite{Widschwendter2018Epigenome-basedChallenges,Stricker2017FromEpigenomics}.
In this section, we survey several problems in the analysis of epigenomic data and some methods designed to solve them.

\subsection{Semi-automated genome annotation}

To get a complete picture of the epigenomic state of each part of the genome, researchers must combine the results of a number of assays.
Large consortia have produced datasets that examine many aspects of epigenomic state~\rev{\cite{Kundaje2015IntegrativeEpigenomes,ENCODEProjectConsortium2004TheProject,Bujold2016ThePortal}}, and one can combine these into a data matrix.
One can divide this data matrix into row vectors, one for each \rev{assay.} 
Alternatively, one can divide the matrix into column vectors, one for each position in the genome.
Either way, the raw signal data proves difficult to interpret and explore on its own.

Semi-automated genomic annotation~(SAGA) methods~\cite{Libbrecht2015JointExpression} aid in this process by clustering regions of the genome by similarity in terms of epigenomic properties.
One might describe the task in terms of identifying clusters of similar column vectors in the data matrix.
However, we cannot assume independence between the column vectors.
In fact, data in each column vector is highly dependent on its neighbors.
Therefore, SAGA methods also simultaneously segment the genome, defining the width of a region dynamically and heterogeneously.
This process results in a partition of the genome called a \emph{segmentation}, with every region assigned to a different cluster, usually called a \emph{label}~\cite{Hoffman2012UnsupervisedSegmentation} or \emph{chromatin state}~\cite{Ernst2012ChromHMM:Characterization}.

We can almost completely automate the simultaneous segmentation and clustering \rev{of} a SAGA method.
The \rev{``semi''} in ``semi-automated genome annotation'' refers to the interpretation of the resulting clusters, conducted by a human expert.
The expert examines both individual segments and aggregate features of each cluster, and describes the captured pattern in terms of a putative biological role.
The identified roles may include the start of a gene, the end of a gene, and an \emph{enhancer}---a kind of genomic element that drives expression of apparently distant genes---as well as many others.
All of these have a characteristic epigenomic pattern, and SAGA methods help to characterize new instances of this pattern~\cite{Hoffman2013IntegrativeData}.
Researchers have used these methods to annotate many genomes, including human~\rev{\cite{Hoffman2012UnsupervisedSegmentation, Ernst2012ChromHMM:Characterization,Day2007UnsupervisedData,Zhang2016JointlyTypes}}, mouse~\cite{Yue2014AGenome}, and fruit fly~\cite{Kharchenko2011ComprehensiveMelanogaster}, enabling researchers to quickly assign function to genomic regions.

\rev{Methods such as HMMSeg~\cite{Day2007UnsupervisedData}, ChromHMM~\cite{Ernst2012ChromHMM:Characterization}, Segway~\cite{Hoffman2012UnsupervisedSegmentation}, EpiCSeg~\cite{Mammana2015ChromatinEpigenome}, and IDEAS~\cite{Zhang2016JointlyTypes}} provide an unsupervised learning approach to finding regions with similar characteristics.
Most of these methods employ graphical models to find similarities in epigenomic data across genomic regions.
These models treat the observed data as being emitted by some theoretical state with defined parameters, reflecting the function of that region.
The first SAGA method, HMMSeg~\cite{Day2007UnsupervisedData}, takes a collection of input epigenomic assays, smooths the data with wavelets, and uses a hidden Markov model~\cite{Rabiner1989ARecognition,Baum1970AChains,Baum1966StatisticalChains,Baum1972AnProcess,AmericanMathematicalSociety.1968PacificMathematics,Blakley1964HomogeneousTransformations} where the hidden state represents cluster membership.
ChromHMM~\cite{Ernst2012ChromHMM:Characterization} uses a hidden Markov model that models input signals as vectors of random Bernoulli variables.
The Bernoulli vectorization binarizes input data into discrete ``on'' or ``off'' categories for each region, based on whether or not the signal in that region exceeds a significance threshold based on a Poisson background distribution.
EpiCSig~\cite{Mammana2015ChromatinEpigenome} uses a similar approach, although it takes raw sequencing counts and models them as emissions from negative binomial distributions instead.
Segway~\cite{Hoffman2012UnsupervisedSegmentation}, conversely, uses single- or multiple-component Gaussians to model real-valued signal data~\cite{Chan2018SegwayTraining}.
Segway generalizes the hidden Markov model with a dynamic Bayesian network~\cite{Dagum1995UncertainForecasting} that can impose hard constraints on segment lengths.
Segway can also perform semi-supervised learning, and an extension enables using it in a fully-supervised pipeline~\cite{Libbrecht2018ATypes}.
\rev{Finally, IDEAS~\cite{Zhang2016JointlyTypes} iteratively} segments the genome for multiple input cell types at once, and classifies similar regions from across cell types using an infinite-state hidden Markov model.

\subsection{Transcription factor binding site prediction}
\emph{Transcription factors} form a class of proteins that bind to chromatin and activate or repress gene expression.
There are over 1,600 likely transcription factors, each with a characteristic pattern of binding in different cell types~\cite{Vaquerizas2009AEvolution,Lambert2018TheFactors}.
Understanding where transcription factors bind, and why, is crucial to a mechanistic understanding of gene regulation.
As transcription factors influence the rate of gene expression, knowing where transcription factors bind can help predict when transcription occurs.
The most widely-used method to determine transcription factor binding in living cells is ChIP-seq~\cite{Johnson2007Genome-wideInteractions}.
The\-se methods sequence protein-bound DNA, determining the positions at which the DNA comes in close proximity to a particular transcription factor.
Related methods such ChIP-exo~\cite{Rhee2012ChIP-exoAccuracy}, ChIP-nexus~\cite{He2015ChIP-nexusFootprints}, and Cleavage Under Targets and Release Using Nuclease (CUT\&RUN)~\cite{Skene2017AnSites} improve on the initial approach.

The existing assays for determining transcription factor binding locations fail under many conditions.
Most of these methods, require an antibody specific to the target of interest, which sometimes cannot be produced.
Other methods, like CETCh-seq~\cite{Savic2015CETCh-seq:Proteins}, require editing the genome in ways that might cause unexpected side effects.
Furthermore, these assays all require more biological material than researchers \rev{can typically obtain from patient samples.}

Computational approaches, however, can predict binding for many transcription factors at once without requiring specific antibodies or large numbers of cells.
These approaches have the goal of predicting a transcription factor's binding at each genomic region.
Several methods tackle prediction by inferring transcription factor occupancy from DNA-binding \emph{motifs}.
These motifs consist of short, recurring DNA sequences to which one transcription factor binds~\cite{Dhaeseleer2006WhatMotifs,Bailey1998CombiningSearches,Grant2011FIMO:Motif,Thomas-Chollier2008RSAT:Tools}.
Most often, we represent a motif as a \emph{position weight matrix}~\cite{Stormo1982UseColi,Wasserman2004AppliedElements} which characterizes the expected frequency of each base's occurrence within a binding sequence.
Motifs can come from ChIP-seq data but often come from simple extracellular experiments such as protein-binding microarrays~\cite{Badis2009DiversityFactors} or HT-SELEX (high throughput systematic evolution of ligands by exponential enrichment)~\cite{Ogawa2012High-ThroughputVitro}.
The MEME method for motif elucidation searches for recurring motifs in a given set of genomic regions using an expectation maximization algorithm~\cite{Bailey1995UnsupervisedMaximization}.
When \rev{gi\-ven} transcription factor binding positions from ChIP-seq data, this reveals recurring motifs for that transcription factor.
\rev{Unfortunately,} predictions that use sequence motifs alone~\cite{Jayaram2016EvaluatingPrediction} do not identify experimentally verifiable binding sites with sufficient utility for genome-wide use.
A pair of observations state this principle: the \emph{futility conjecture}~\cite{Wasserman2004AppliedElements} and the \emph{dual futility conjecture}~\cite{Karimzadeh2018VirtualTranscriptome} \rev{(see \autoref{tab:epig-glossary-assays})}.

To move beyond the futility of predicting transcription factor binding sites with sequence alone, most methods integrate additional data.
Sometimes these data include other epigenomic data, such as chromatin accessibility data, that either already exist in public databases or that one can obtain much more easily than a new ChIP-seq assay.
CENTIPEDE~\cite{Pique-Regi2011AccurateData} predicts binding sites using a transcription factor's position weight matrix along with open chromatin or histone modification epigenomic data.
It first finds all regions which match a known sequence motif, then uses the shape of signal in other epigenomic assays to cluster each match.
CENTIPEDE calculates the posterior probability that a transcription factor binds a genomic region given other information from other epigenomic assays.
For instance, a transcription factor bound to DNA will leave an inaccessible region in chromatin accessibility data.
Since chromatin accessibility assays mark regions with bound transcription factors as inaccessible, searching for these inaccessible regions can inform whether or not a transcription factor is bound.
HINT~\cite{Gusmao2014DetectionModifications} searches for the same patterns in chromatin accessibility and histone modifications, but delineates regions by detecting sudden changes in epigenomic signal.
By modeling ChIP-seq data from histone modifications and an input chromatin accessibility experiment using a hidden Markov model, HINT can \rev{find} transcription factor binding without motif information.
It can also incorporate transcription factor motifs and rank them.
Methylphet~\cite{Xu2015Base-resolutionVivo} incorporates \emph{DNA methylation} information, training a random forest on bisulfite sequencing~(BS-seq) data and ChIP-seq on one transcription factor.
This random forest can then predict transcription factor binding sites using only BS-seq data on another sample.

Other methods use increasing numbers of data types to predict transcription factor binding sites.
FactorNet~\cite{Quang2017FactorNet:Data} applies a deep neural network to this problem.
FactorNet trains on input DNA sequences, chromatin accessibility, gene expression, and the binding status of a given transcription factor.
It uses this network to predict the binding status of new input sequences, chromatin accessibility, and expression levels.
Keilwagen et al.~\cite{Keilwagen2017LearningBinding} combine features from both previous genomic annotations, \emph{de novo} motifs from ChIP-seq and DNase-seq, and raw sequence-level data including RNA-seq.
They model each of these features in a different manner.
Gaussians model numerical features like RNA-seq expression levels, binomial distributions model discrete features like gene annotations, and they use a third order Markov model for genomic sequence.
For a new cell type, they then take \rev{an} average of the prediction scores from these models to obtain a new prediction of transcription factor occupancy.
This algorithm tied for best performance in the ENCODE-DREAM \emph{in vivo} Transcription Factor Binding Site Prediction Challenge~\cite{2017ENCODE-DREAMSyn6131484}.
Virtual ChIP-seq~\cite{Karimzadeh2018VirtualTranscriptome} deemphasizes motifs, relying more on open chromatin data and ChIP-seq data from other cell types~\cite{Karimzadeh2018VirtualTranscriptome}.
It also uses data from RNA-seq, a method for determining steady-state gene expression.
Virtual ChIP-seq uses a multi-layer perceptron to integrate these diverse data types and others, learning different hyperparameters and weights for each transcription factor.

\subsection{Topologically associated domain prediction}

While computational biologists usually represent the ge\-no\-me as a simple string of letters, it actually has a complex three-dimensional structure.
Beyond the fine-scale structure inherent in nucleosome \rev{positioning,} each chromosome in a cell's nucleus has higher-order structures that persist in 3D. 
These structures bring together regions of the genome distant in one dimension, resulting in long-range chromatin interactions between genes and enhancers.

Chromosome conformation capture (3C) assays quantify sp\-atial proximity between specific genomic regions.
Some of these assays, such as Hi-C~\cite{Lieberman-Aiden2009ComprehensiveGenome} and ChIA-PET~\cite{Fullwood2009AnInteractome}, interrogate spatial proximity in a whole-genome all-versus-all fashion.
Another recent technique, Hi-ChIP~\cite{Mumbach2016HiChIP:Architecture}, combines methods from ChIP-seq to only find large regions nearby a protein of interest.
These techniques have found self-interact\-ing regions at various scales that are \emph{conserved} across species~\cite{Dixon2012TopologicalInteractions}.
\emph{Topologically associated domains}~(TADs) are persistent structures of spatial proximity approximately 1\,Mbp in length~\cite{Dixon2012TopologicalInteractions,Rao2014ALooping}.
Rather than producing a vector like other epigenomic sequencing assays, these techniques produce a triangular matrix of each potential interaction.
Unfortunately, as the number of potential interactions grows with the square of the number of regions interrogated, the sequencing necessary to produce it becomes rather expensive.

Many methods predict TAD locations from Hi-C data, such as Chrom3D~\cite{Paulsen2017Chrom3D:Contacts} and TADbit~\cite{Serra2017AutomaticColors}.
These tools use 3C-class data to get the proximity of genomic regions to each other, and use this information to infer TAD positioning.
Chrom3D~\cite{Paulsen2017Chrom3D:Contacts} uses a Monte Carlo simulation to model histones as beads-on-a-string.
Its Monte Carlo simulation minimizes a loss-score function \rev{with} Hi-C and ChIP-seq data as input.
The final output includes both a visualization of the chromatin, and the position of the identified TADs.
TADbit~\cite{Serra2017AutomaticColors} uses a breakpoint detection method to segment the genome by finding the optimal balance between the amount of Hi-C interactions upstream, downstream, and within TADs.
An optimal segmentation will maximize the total log-likelihood such that all three interaction categories are equal.

Rao et al.~\cite{Rao2014ALooping} have shown that chromatin compartmentalizes itself into either gene dense, highly expressed regions, or lowly expressed regions.
They used a Gaussian hidden Markov model on Hi-C interaction data to find large-scale \rev{self-interact\-ing} regions, and inferred compartmentalization from this.
Methods like BACH-MIX~\cite{Hu2013BayesianChromosomes}, and MEGABASE~\cite{DiPierro2017DeArchitecture}, have been developed to determine which compartment each genomic region belongs to.
BACH-MIX uses Markov chain Monte Carlo techniques to converge on a 3D model of chromatin that agrees with experimental 3C-class data.
Since this experimental data can assay a heterogeneous population, where chromatin can freely move between multiple states, BACH-MIX takes into account multiple spatial rearrangements simultaneously.
It models each genomic region as two substructures whose spatial arrangement varies in the sample assayed.
By modeling the uncertainty between the possible arrangements with a mixture component model, it reconstructs likely chromatin architectures and their compartmentalization.
MEGABASE predicts structure without 3C-class data, instead determining chromatin compartmentalization from histone modifications.
It models DNA as a polymer of self-inter\-act\-ing loci based on ChIP-seq data, and trains a neural network to predict compartmentalization based on this model.

\subsection{Histone modification and DNA methylation prediction}

Histone modification prediction also benefits from computational alternatives to ChIP-seq.
Epigram~\cite{Whitaker2015PredictingMotifs} identifies sequence motifs across cell types that strongly hint at histone modifications.
Epigram then employs a random forest classifier to predict histone modification and DNA methylation from these motifs.
ChromImpute~\cite{Ernst2015Large-scaleTissues} predicts, from a core set of commonly performed epigenomic assays, signal from other epigenomic assays.
To do this, ChromImpute trains regression trees on samples where the data type of interest exists.
By averaging the results of the trees from these previous experiments, ChromImpute infers signal from unperformed experiments.
PREDICTD~\cite{Durham2018PREDICTDDecomposition} imputes missing histone modification and methylation signals with large factor decomposition.

\section{Noncoding variant effects}\label{sec:noncoding}

Researchers and medical professionals often want to know what effects DNA changes will have on cellular and organismal phenotype.
While interpreting the effects of changes to the sequence coding for proteins is relatively easy, interpreting the \emph{noncoding} sequence that makes up most of a complex organism's genome has proven far more challenging.
Many noncoding sequence variants are associated with particular phenotypic traits or genetic diseases~\cite{Hindorff2009PotentialTraits}.
Noncoding changes often cause phenotypic effects mediated through epigenomic and gene expression changes~\cite{Prensner2011TheBiology}.
We wish to distinguish benign noncoding variants from those that are \emph{deleterious}.
Deleterious noncoding effects often occur in specific regions that control gene regulation, called \emph{regulatory regions} as a class.
Regulatory regions include enhancers~\cite{Shlyueva2014TranscriptionalPredictions} and regions at the start of a gene~\cite{Riethoven2010RegulatoryInsulators}.

Some methods aim to identify regulatory regions and deleterious noncoding changes based on sequence alone.
For example, gkm-SVM~\cite{Ghandi2016GkmSVM:SVM,Ghandi2014EnhancedFeatures} \rev{finds} short sequences (k-mers) that are indicative of enhancer activity.
It then uses a support vector machine~(SVM) to find enriched k-mers in the training set versus a background of random sequences.
It also allows these k-mers to have an arbitrary number of breaks, or gaps, in the sequence.
The training dataset generally consists of binding sites for a given transcription factor.
The kernel for this SVM computes a similarity score between two sequences, which are represented as short sequences including gaps.
DeepSEA~\cite{Zhou2015PredictingModel} trains a deep convolutional neural network on genomic sequence to predict epigenomic state.
It can predict both transcription factor binding and histone modification status.
DeepSEA examines the impact of sequence changes by comparing predictions made for both unmodified and modified sequence.
Basset~\cite{Kelley2016Basset:Networks} learns chromatin accessibility from sequence alone.
It uses a deep convolutional neural network on the sequence to obtain probability predictions of DNase-seq signal.

We can also determine a mutation's deleteriousness by integrating genomic \emph{conservation} data.
Conservation measures how little a sequence has changed over the course of evolution.
Mutations almost certainly have occurred in conserved regions over evolutionary time, but those that decrease organismal fitness will have greatly diminished prevalence today.
We therefore assume that sequences that remain conserved across species or among populations in the same species indicate that mutations there would be highly deleterious, cause disease\rev{, or cause} death.

Several methods use conservation to identify deleterious mutations.
Combined Annotation Dependent Depletion~(CADD) integrates 63 features, including annotations drawn from conservation and epigenomic data, using a linear kernel SVM~\cite{Kircher2014AVariants}.
To label the SVM's training data, the CADD authors distinguish between common sequence variants that have changed since the human--chimpanzee common ancestor, and depleted simulated variants.
Eigen, by contrast, applies an unsupervised method that uses conservation scores, protein function scores, and allele frequencies from a variety of mutation databases~\cite{Ionita-Laza2016AVariants}.
By combining these into a block matrix, and taking the eigendecomposition of that matrix, Eigen finds each mutation's predictive accuracy for deleteriousness.

Some methods for predicting deleterious noncoding \rev{sequen\-ce} variants rely on Inference of Natural Selection from Interspersed Genomically coherent elements~(INSIGHT)~\cite{Gronau2013InferenceDivergence} to identify the strength of natural selection on these variants.
INSIGHT uses a complex evolutionary model that incorporates knowledge from multiple species and accounts for heterogeneous observations at different parts of the genome.
The fitCons method clusters DNase-seq, RNA-seq, and histone modification data not unlike the SAGA methods above~\cite{Gulko2015AGenome}.
It then estimates the fraction of bases within each cluster that INSIGHT identifies as strongly under natural selection.
fitCons labels each genomic region with an importance score based on INSIGHT's natural selection probability.
LINSIGHT uses mostly the same procedure as fitCons, but eschews fitCons' clustering step for a generalized linear model relating observed epigenomic features to INSIGHT scores~\cite{Huang2017FastData}.
Like fitCons, it outputs INSIGHT-scored fitness for each genomic region.

\section{Integrative single-cell analysis}
\label{sec:singlecell}

\begin{figure}[t]
\centering
\includegraphics[width=\linewidth]{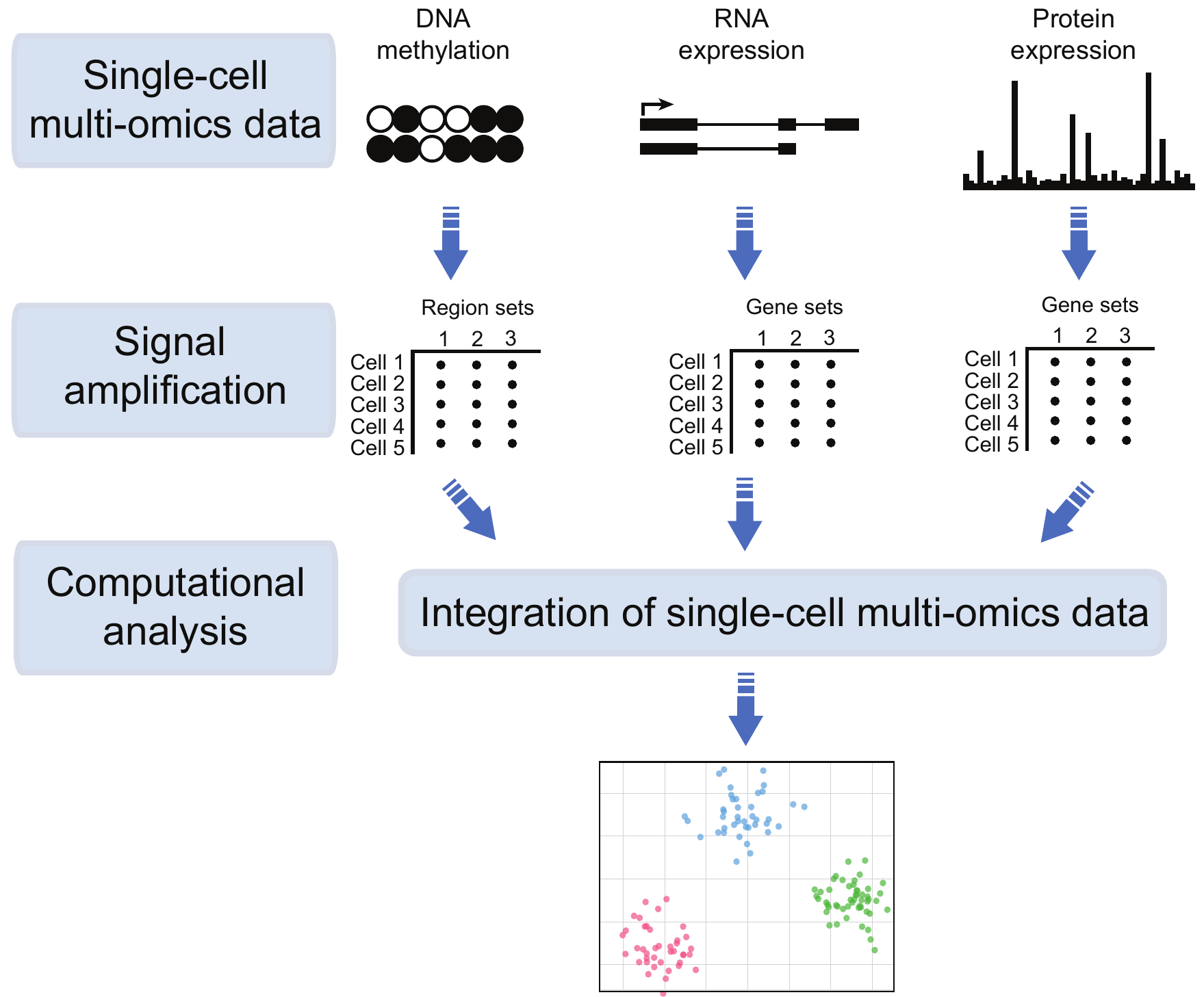}
\caption{\textbf{\rev{Single-cell multi-omics data integration.}} \rev{A typical single-cell data analysis consists of} three steps. First, the raw data are preprocessed, filtered, and quality-controlled separately for each assayed omics dimension, accounting for \rev{analytical challenges}, such as technical variation, sparse signal, and amplified artifacts. Second, as single-cell data are intrinsically of low coverage, it is a good practice to increase the \rev{signal-to-noise} ratio by aggregating data; for example, by combining expression levels of genes \rev{with similar functions} or similar DNA methylation levels across genomic regions bound by the same transcription factors. Finally, data are integrated into one multi-omics \rev{map.}}
\label{fig:multiomics}
\end{figure}

A major question in biology is how to describe and quantify every cell in a multicellular organism~\cite{regev2017science}, such as human, \rev{which can contain} a myriad of different types of cells.
\rev{{\em Cell type}, such as muscle or nerve, is typically defined based on function of a tissue in which the cells reside and unique morphological properties of that tissue}~\cite{Clevers2017your}.
However, considerable cell-to-cell variation in cells within a single cell type \rev{indicates the existence of distinct} cell states (\eg, mitotic, \rev{migratory}) and various cell \rev{behaviors, which depend} on local activity of each cell in a particular microenviroment.
Even within a single tissue, there are diverse populations of cells, representing different manifestations of that tissue.

A traditional approach to studying tissues \rev{relies} on a pooled assay and \rev{uses} a weighted average of a \rev{{\em bulk sample of cells}} from a particular tissue (\ie, a large population of cells), \rev{which can obscure variations across cells within the sample}.
Advances in single-cell technologies have enabled measurements at {\em single-cell resolution} and have opened new avenues to investigate the heterogeneity of cells across tissues and within cell populations~\cite{kelsey2017single}.
Single-cell technologies \rev{profile} individual cells from various perspectives, including genomic~\cite{gawad2016single}, epigeno\-mic~\cite{schwartzman2015single}, transcriptomic~\cite{stegle2015computational}, and proteomic~\cite{wu2012single} \rev{perspectives}.
However, multi-omics single-cell measurements pose a significant challenge for data analysis, integration, and interpretation~\cite{yuan2017challenges}, one that could benefit from machine learning.
Integrative single-cell analyses focus on: (1) identification and characterization of cell types and the study of their organization in space and over time, and (2) inference of gene regulatory networks from multi-omics data and assessment of network robustness across cells.

\subsection{Cell type discovery and exploration}

Single-cell RNA sequencing (scRNA-seq) is a powerful technology to measure \rev{gene expression of individual cells and to characterize} heterogeneity and functional diversity of cell populations~\cite{shapiro2013single}. \rev{To characterize a cell population one needs to identify what genes are expressed in each cell and how strong that gene expression is.} \rev{Information on heterogeneity of cells in a given sample} can answer questions \rev{that cannot be addressed} by traditional ensemble-based met\-hods, \rev{in which} gene expression measurements are averaged over \rev{all cells} pooled together.  Recent studies have demonstrated that {\em de novo} cell type discovery and identification of functionally distinct cell subpopulations are possible via unbiased analysis of all transcriptomic information provided by scRNA-seq data~\cite{poirion2016single}. However, compared with bulk RNA-seq data, unique challenges associated with scRNA-seq include high dropout rate~\cite{zheng2017massively} (where a large number of genes have zero reads in some cells, but relatively high expression in the remaining cells) and \rev{curse of dimensionality} (\rev{where distances between cells become more and more similar as dimension of the space they reside within increases}, \eg, \rev{mammalian} expression profiles are frequently studied as vectors with about 20,000 genes)~\cite{yuan2017challenges}. 

\rev{To} address these challenges, various unsupervised computational algorithms~\cite{wang2017visualization, pierson2015zifa, cleary2017efficient,kiselev2017sc3,butler2018integrating} have been proposed since the first study of scRNA-seq~\cite{tang2009mrna}. Most of these computational algorithms either rely on dimension-reduction techniques~\cite{pierson2015zifa,cleary2017efficient,butler2018integrating} or utilize a \rev{consensus} from multiple clustering results~\cite{wang2017visualization, kiselev2017sc3}. For example, Zero Inflated Factor Analysis (ZIFA), one of the very first \rev{dimensionality reduction} methods to address the dropout events,
assumes that the dropout rate for a gene follows a double exponential distribution with respect to the expected expression level of that gene in the population~\cite{pierson2015zifa}. CellTree~\cite{yotsukura2016celltree} incorporates a Latent Dirichlet Allocation model with latent gene groups to measure \rev{cell-cell} distance by a detected tree structure outlining the hierarchical relationship between single-cell samples to introduce biological prior knowledge. \rev{Cleary et al.~\cite{cleary2017efficient} take} another perspective by utilizing compressed sensing together with \rev{the assumption} that scRNA-seq data \rev{are collected} in a compressed format, as composite measurements of linear combinations of genes. However, \rev{a clear disadvantage} of these \rev{dimensionality reduction methods} is \rev{a strong statistical assumption of data following an appropriate distribution.} \rev{Such assumptions do not always} hold \rev{and depend on a particular scRNA-seq technology or platform.}

\rev{Different} from \rev{dimensionality reduction} methods, \rev{ensemble methods first} generate multiple approximate representations or clusterings for cells and then integrate them in a principled way. For instance, SIMLR~\cite{wang2017visualization} first generates multiple kernels to represent approximate \rev{cell-cell} variabilities and then uses a non-convex optimization framework to refine and integrate these kernels and output a detailed and fine-grained description of \rev{cell-cell} similarity matrix. This learned similarity matrix can enable efficient clustering and visualization for scRNA-seq data. SC3~\cite{kiselev2017sc3} takes a similar strategy in that it first generates multiple clustering results with different subsets of genes and then combine these clustering results with majority voting. 

\rev{The scRNA-seq data analysis methods described so far} deal with scRNA-seq data generated by a single experiment. When it comes to integrative analysis of scRNA-seq data from multiple patient groups, different samples across tissues, and multiple conditions, the number of available methods is limited. The unique challenge lies in the fact that the accompanying biological and technical variation tends to dominate the signals for clustering the pooled single cells from the multiple populations. A recent effort~\cite{Zhang2018AMC} developed a multi-task clustering method to address the problem. This method introduces a multi-task learning method with embedded feature selection to simultaneously capture the differentially expressed genes among cell clusters and across all cell populations or experiments to achieve better single-cell clustering accuracy.

\subsection{Single-cell multi-omics analysis}

Beyond \rev{scRNA-seq data,} other single-cell sequencing techniques measure various biological dimensions, such as DNA methylation~\cite{smallwood2014single}, histone modifications~\cite{rotem2015single}, open chromatin (scATAC-seq and scDNase-seq~\cite{buenrostro2015single,cusanovich2015multiplex}), chromosomal conformation~\cite{nagano2013single}, proteome~\cite{frei2016highly}, and metabolome~\cite{fessenden2016metabolomics}.
\rev{Sin\-gle-cell} multi-omics data are potentially more powerful to provide a comprehensive understanding of the cells than any single-omics data~\cite{Macaulay2017single}, but their analysis poses interesting challenges for machine learning.
In particular, one needs to discover not only information shared across various omics data but also complementary signals that are specific to a particular omics data type (\autoref{fig:multiomics}).

\rev{Current} methods for analysis of single-cell multi-omics data are correlation-based or clustering-based~\cite{bock2016multi}.
First, a prevailing approach considers pairs of omics \rev{datasets} and generates hypotheses by measuring correlations \rev{between datasets}. For example, several studies~\cite{angermueller2016parallel,hou2016single,macaulay2015g,han2018sidr} apply canonical correlation analysis (CCA)~\cite{witten2009penalized,waaijenborg2008quantifying,le2009sparse}, a method that has been widely used in bulk data \rev{analysis,} to estimate correlations between single-cell DNA methylation and \rev{scRNA-seq data}.
CCA learns a low-dimen\-sio\-nal representation of \rev{omics datasets} that captures common information shared across all \rev{datasets}.
However, the CCA-based analysis is limited \rev{because it cannot take} into account {\em dropout events}.
Dropout events are a special type of missing values caused by the low number of RNA transcriptomes in the sequencing experiment and the stochastic nature of gene expression at a \rev{single-cell} level.
Consequently, these dropout events become zeros in a gene-cell expression matrix and these ``false zeros'' mix with ``true zeros'' representing genes not expressed in a cell at all.
To conquer this dropout issue, imputation methods use correlations between multi-omics data to impute the missing values.
For example, MAGIC~\cite{van2017magic} imputes the missing values by applying a diffusion model to gene-gene correlation matrix. \rev{Similarly, scImpute~\cite{li2018accurate} pulls} information from groups of similar cells to \rev{complete a sparse data matrix and obtain a better representation of cell-cell correlations.}

Another direction for integrating single-cell multi-omics da\-ta uses a two-stage approach: first, construct a separate clustering for each omics dataset, and then combine these clusterings for comparison and analysis~\cite{macaulay2015g, cheow2016single,peterson2017multiplexed, stoeckius2017simultaneous}.
The advantage of such an approach is its ability to infer importance of each data type and to identify information common to all data types.
For example, studies~\cite{peterson2017multiplexed,stoeckius2017simultaneous} adopt the method that first clusters cells based on each \rev{omics dataset} and then perform extensive comparisons between clusters using statistical association tests.
Along similar lines, MATCHER~\cite{welch2017matcher} uses manifold alignment of single-cell multi-omics data.
MATCHER first uses a Gaussian process latent variable model to independently cluster every cell in each \rev{omics dataset}.
It then aligns these clusterings and combines them into a global clustering of cells.
These clustering approaches have the advantage of detecting both complementary and common patterns in single-cell multi-omics data.
Nevertheless, they can suffer from computational complications caused by extensive generation and statistical comparison of many clusterings.

\subsection{Large-scale single-cell bioinformatics}

As single-cell technologies advance, the number of cells generated per each experiment increases, demanding for efficient and large-scale bioinformatics~\cite{zheng2017massively}.
Present approaches for large single-cell data utilize: (1) approximate inference~\cite{iacono2018bigscale} and fast software implementations~\cite{wolf2018scanpy}, or (2) adopt deep learning methods that take small batches of cells as input~\cite{lin2017using,amodio2017exploring}.
For example, bigScale~\cite{iacono2018bigscale} uses large sample sizes to approximate an accurate numerical model of noise and cluster datasets with millions of cells.
SCANPY~\cite{wolf2018scanpy}, however, provides an efficient Python-based implementation that is easy to interface with other machine learning packages, such as Tensorflow~\cite{abadi2016tensorflow}.

\rev{Another} direction within this vein is to use deep-learning based methods, since they can naturally train a multi-layer neural network using memory-efficient mini-batch stochastic gradient descent.
For example, \cite{lin2017using} apply deep auto-encoders to obtain low-dimensional representations that optimize the reconstruction of original noisy inputs.
Similarly, SAUCIE (Sparse Autoencoder for Unsupervised Clustering, Imputation, and Embedding)~\cite{amodio2017exploring} uses a multi-task deep auto-encoder and performs several key tasks for single-cell data analysis including clustering, batch correction, visualization, denoising, and imputation.
SAUCIE is trained to reconstruct its own input after reducing its dimensionality in a 2D embedding layer, which can be used to visualize the data.
Different from traditional deep auto-encoders, SAUCIE uses two additional model regularizations: (1) an information dimension regularization to penalize the entropy as computed on the normalized activation values of each neural layer and thereby encourage binary-like encodings amenable to clustering, and (2) a maximal mean discrepancy (MMD) penalty to correct for batch effects.
Although these deep learning methods achieve promising results and are capable \rev{of dealing} with large single-cell data, their black-box nature and lack of interpretability limit their wide adoption in practice.

\section{Cellular phenotype and function}\label{sec:function}

\begin{figure}[t]
\centering
\includegraphics[width=\linewidth]{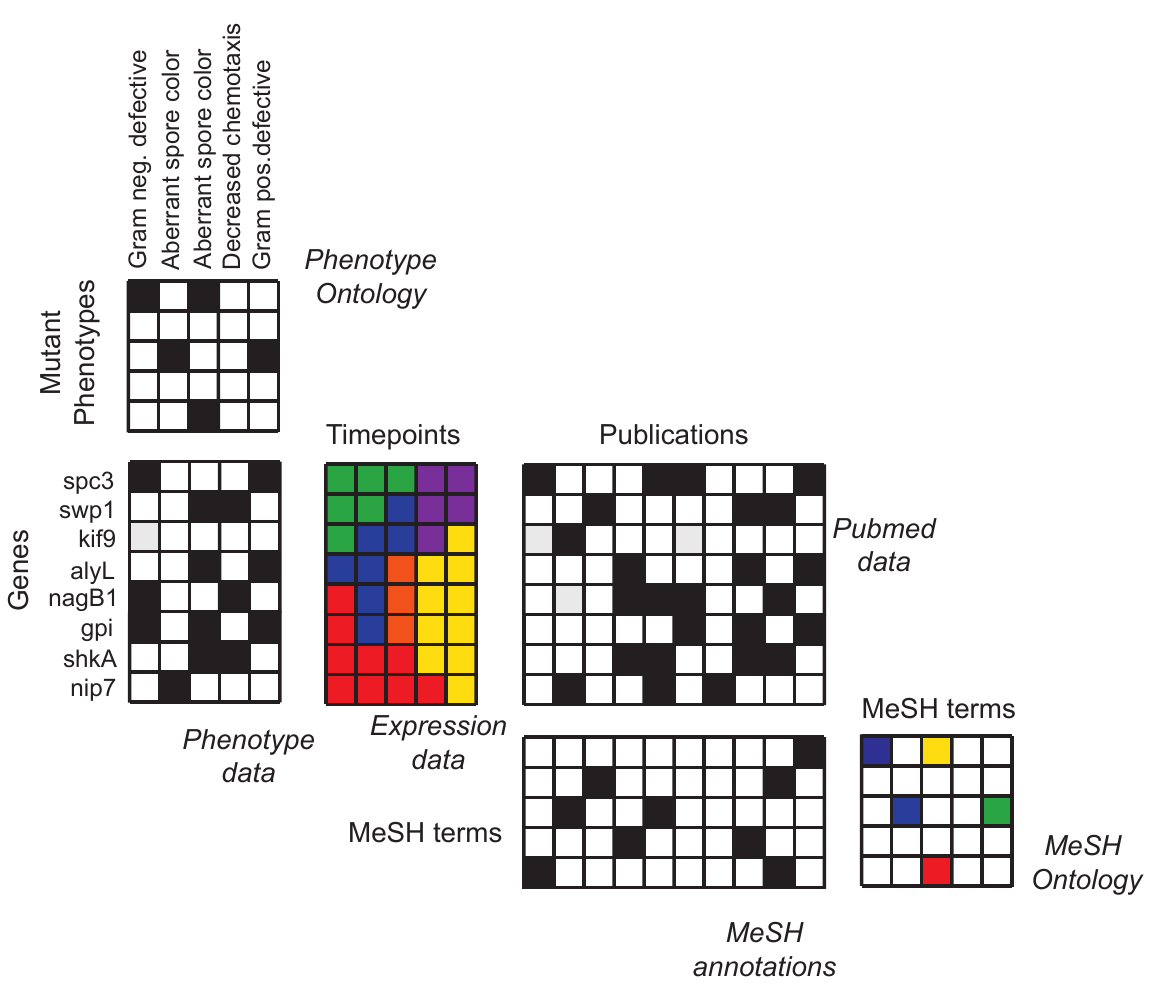}
\caption{\textbf{A matrix-based representation of diverse datasets relevant for gene function prediction.} Let us consider a hypothetical gene function prediction task. Here, the function is {\em response to bacterial infection}~\cite{Typas2015bacterial}, meaning that the task is to identify genes in an eukaryotic organism that determine how the organism will respond to a bacterial infection. There is a variety of diverse datasets potentially relevant for this task and each dataset is typically represented with a separate data matrix. Shown is an example with six data matrices, including gene-phenotype associations, gene expression profiles, biomedical literature, and annotations of research papers. Integrative approaches solve the gene function predict task by establishing a rigorous statistical correspondence between different input dimensions of these seemingly disparate data matrices~\rev{\cite{Zitnik2016jumping,Zitnik2015data,Wang2014similarity,Gonen2012predicting,Gligorijevic2014integration,Zitnik2014drug,Zitnik2014matrix,Gligorijevic2015fuse,Strazar2016orthogonal,Gligorijevic2016patient}}. For example, genes can be linked to \rev{Medical Subject Headings (MeSH concepts)} via gene-publication relationships (\ie, lists of genes discussed in a given research paper), followed by publication-MeSH relationships (\ie, lists of the MeSH concepts assigned to a given research paper). For example, a collective matrix factorization approach in \cite{Zitnik2015data} can fuse such complex systems of data matrices. The approach has been used to predict gene functions in various species~\cite{Zitnik2015data,Zitnik2014matrix} and has subsequently been applied to prioritization of genes mediating bacterial infections~\cite{Zitnik2015collage}.}
\label{fig:function-prediction}
\end{figure}

Our ability to generate sequence data has been improving at a rapid rate for the past decade, and this trend is likely to continue for the next decade (Section~\ref{sec:epigenomic}).
A vast majority of these sequences are of proteins of unknown function and their worth could be substantially increased by knowing the biological roles that they play.
Accurate annotation of protein function is a key to understanding life at the molecular level and has great biomedical and pharmaceutical implications.
To this aim, numerous research efforts, such as the Encyclopedia of DNA Elements (ENCODE) \citep{Encode2012integrated} (Section~\ref{sec:epigenomic}) and the Genotype-Tissue Expression (GTEx)~\citep{Gtex2015genotype} \rev{projects,} have expanded the breadth of available data that lend themselves to protein function prediction (\autoref{fig:function-prediction}).

{\em Protein function} is a concept describing biochemical and cellular aspects of molecular events that involve proteins.
Protein functions can be divided into three major categories: (1) molecular functions, \eg, the specific reaction catalyzed by an enzyme, (2) biological processes, \eg, the metabolic pathway the enzyme is involved in, and (3) systems or physiological events, \eg, if the enzyme is involved in respiration, photosynthesis or cell signaling.
One can also consider a fourth level, \ie, cellular components, describing cell compartments in which proteins have a role, such as a cell membrane and organelles.
Functions of proteins can also vary in space and time as in the case of moonlighting proteins (\rev{\ie, multitask} proteins).
Furthermore, many protein functions are carried out by groups of interacting proteins and these interactions can be predicted.

Most proteins are poorly characterized experimentally and we know little about their functions.
Furthermore, vast majority of proteins with known functions are from model organisms, but even for those organisms, a significant part of all proteins coded in their genomes remain to be characterized.
For example, in {\em Escherichia coli}, about one third of the 4,225 proteins remain functionally unannotated (\ie, orphan proteins) and a similar proportion applies to \rev{{\em Saccharomyces cerevisiae}.}

\subsection{Protein function prediction}\label{sec:protein-function-prediction}

Protein functions can be inferred on the basis of amino acid sequence similarity~\cite{Zhang2017cofactor}, gene expression~\cite{Wan2017analysis}, protein-pro\-tein interactions~\cite{Mostafavi2008genemania,Amar2014constructing,Zhang2017cofactor}, meta\-bolic interactions~\cite{Manichaikul2009metabolic}, genetic interactions~\cite{Kuzmin2018systematic}, evolutionary relationships~\cite{Gaudet2011phylogenetic}, 3D structural information~\cite{Konc2014binding}, mining of \rev{biomedical text}~\cite{You2017deeptext2go}, and any combination of these data.
At the most basic level, protein function prediction methods can be categorized into two categories: (1) unsupervised similarity-based methods using a principle that similar proteins share similar functions, and (2) supervised methods using a classification of protein functions in the \rev{Gene Ontology (GO)}~\cite{Ashburner2000gene}.

Similarity-based prediction methods relate a functionally uncharacterized protein with proteins whose functions are already known.
The simplest and most often used approach uses sequence similarity search.
Given a query protein, similarity search programs, such as Basic Local Alignment Search Tool \rev{(BLAST)~\cite{Altschul1990basic}, scan} the sequence data banks for homologous proteins of known function or structure and \rev{transfer} their functions to the query protein.
If the query protein is not homologous to any protein with known function, it is possible to {\em de novo} predict functions of the query protein.
A {\em de novo} prediction uses diverse information about the query protein to identify biological properties that are shared among all proteins with the same function (\eg, proteins with the same function might act similarly in similar conditions, for example, in a particular human tissue).
These properties are then used to select proteins whose functions are transferred to the query protein~\cite{Carreras2017comprehensive}.
For example, \cite{Zitnik2015impute,Cho2016compact} developed a low-dimensional matrix decomposition approach that combined genetic interaction networks with other types of gene-gene similarity networks.
These approaches used networks to learn an embedding (\ie, a feature vector) for every protein.
This was accomplished by optimizing a network reconstruction objective, assuming that each protein's embedding depended on embeddings of protein's neighbors in the network.
The learned embeddings were then used as input to \rev{a clustering} algorithm.
Many matrix decomposition~\cite{Zitnik2012nimfa} and tensor factorization~\cite{Nickel2011} methods have proven useful for protein function prediction~\cite{Radivojac2013large}.
For example, \cite{Li2011integrative,Ou2016two} used tensor computations to combine many weighted co-expression gene similarity networks.
The same approach was also used to identify protein complexes, \ie, groups of two or more proteins that form a molecular machinery and together perform a particular function~\cite{Bugge2016combined,Shi2015strategy}.
Along similar lines, \cite{Myers2005discovery,Ray2014bayesian,Greene2015understanding} used Bayesian latent factor models and combined gene expression, copy number variation (CNV), and methylation data to predict protein functions.
As a final example, many approaches aim to understand protein functions by combining data from different tissues~\cite{Greene2015understanding,Ori2015integrated,Zitnik2017ohmnet,Andrews2017cross} or different species~\cite{Deng2010investigating,Hooghe2012flexible,Setty2012inferring,Penfold2015inferring,Imam2015integrated,Ihekwaba2016integrative}.
For example, OhmNet~\cite{Zitnik2017ohmnet} organizes 107 human tissues in a multi-layer network, in which each layer represents a tissue-specific protein-protein interaction network.
OhmNet models the dependencies between network layers (\ie, tissues) using a tissue hierarchy and develops an unsupervised feature learning method then learns an embedding for every node (\ie, protein) in the multi-layer network by considering \rev{edges within each layer (\ie, pro\-tein-protein interactions) as well as edges across layers (\ie, tissue-tissue similarities)}.

\begin{figure}[t]
\centering
\includegraphics[width=\linewidth]{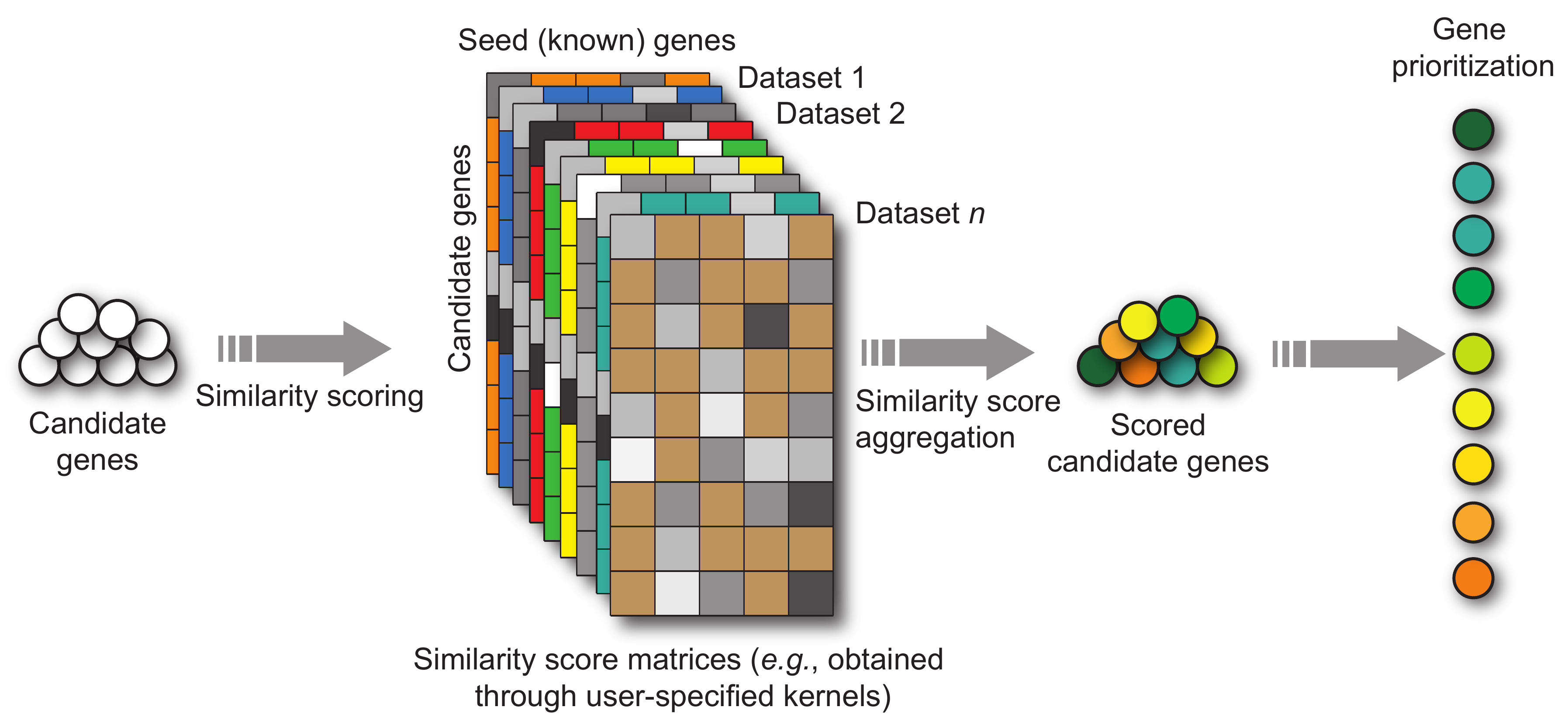}
\caption{\textbf{Gene prioritization.} Gene prioritization aims to identify the most promising genes among a list of candidate genes with respect to a biological process of interest. The biological process of interest is most often represented by a small set of seed genes that are known to be involved in the process. Typically, gene lists generated by traditional disease gene hunting techniques generate dozens or hundreds of genes among which only one or a few are of primary interest. The overall goal is to identify these genes and, in a second step, experimentally validate these genes only. Many different computational methods that use different algorithms, datasets, and strategies have been developed~\rev{\cite{Zitnik2015collage,Sifrim2013extasy,Aerts2006gene,De2007kernel,Chen2009toppgene,Robinson2014improved,Simoes2015neri,Himmelstein2015heterogeneous,Kumar2018pbrit}}. Some of these approaches have been implemented as publicly available tools and several of these approaches have been experimentally validated~\rev{\cite{Zitnik2015collage,Aerts2006gene,Chen2009toppgene,Robinson2014improved,Tranchevent2016candidate}.}}
\label{fig:gene-prioritization}
\end{figure}

If there are examples of proteins with a particular function, they can be used to identify additional proteins with the same function.
This is accomplished by {\em gene prioritization} (\autoref{fig:gene-prioritization}).
Given a set of genes with unknown function, gene prioritization ranks them by their similarity to genes with known function (\ie, seed genes).
Genes at the top of the ranked list are most similar to seed genes and thus are likely to have the same function as seed genes.
Gene prioritization methods can be categorized into four groups: (1) similarity scoring methods that use filtering techniques to independently analyze each dataset~\cite{Franke2004team}, (2) methods that aggregate gene feature vectors from different datasets, \eg, by concatenation, and then use the aggregated vectors as input to a downstream classifier~\cite{Sifrim2013extasy}, (3) methods that use each dataset separately to estimate the similarity of genes with seed genes and then combine similarity scores via a linear or nonlinear weighting~\cite{Lanckriet2004statistical,Aerts2006gene,Tranchevent2016candidate}, and (4) methods that construct a separate gene-gene correlation network for each dataset and combine the networks under supervision of seed genes~\cite{Kohler2008walking,Mostafavi2008genemania}.

Supervised methods for function prediction use a classification of protein functions in \rev{the GO~\cite{Ashburner2000gene}} to specify a supervised prediction task.
\rev{The task presents} four interesting challenges for machine learning methods.
First, functions of proteins are classified into over 40,000 \rev{GO terms,} and this large and complex space represents a challenge for any classification method.
Second, there are dependencies between \rev{GO terms that} lead to situations, where proteins are assigned to multiple functions in the GO, at different levels of abstraction (\eg, {\em cellular transport} versus {\em extracellular amino acid transport}).
Furthermore, proteins typically have multiple different functions, making the function prediction inherently a multi-label, multi-class problem.
Finally, high-level physiological functions, such as {\em inter-cellular transport} or {\em regulation of heart rate}, go beyond simple molecular interactions and require many proteins to participate, and thus such functions usually cannot be predicted by considering a single protein in isolation.
To take on these challenges, many approaches use joint latent factor models~\cite{Zitnik2014matrix,Gligorijevic2014integration}, multi-label learning~\cite{Mostafavi2008genemania}, and ensemble learning~\cite{Wu2010prediction,Pandey2010integrative,Hooghe2012flexible,Bonnet2015integrative}.
A number of machine learning methods \rev{were} also developed to integrate regulatory networks and pathway information to predict functional modules, \ie groups of functionally related proteins~\cite{Heiser2009integrated,Nibbe2010integrative,Zitnik2015netinf,Bonnet2015integrative,Rudolph2016elucidation,Piccolo2016integrative}, which only implicitly invoke the similarity principle described above.

Another consideration is a direct inference of a functional ontology (\ie, a hierarchy of protein functions) from data~\cite{Dutkowski2013gene,Mungall2012uberon}.
For example, \cite{Dutkowski2013gene} use a hierarchical network community detection algorithm together with protein-protein interaction network of {\em Saccharomyces cerevisiae} to infer an ontology whose coverage is comparable to \rev{manually-curated GO annotations.}
Another common approach is to use neural networks to predict protein functions.
For example, \cite{Zitnik2017ohmnet} use a neural network to predict tissue-specific protein functions, \ie, functions taking place in a particular cell type, tissue, organ, or organ system.
Another example that employs neural networks is \cite{Kulmanov2017deepgo}, who use deep learning to learn protein embeddings using protein sequence data, cross-species protein-protein interaction network, and the \rev{GO hierarchical relationships between protein functions.}
Along similar lines, \cite{Ma2018using} use several million genotypes to train a neural network whose architecture is determined by the \rev{GO hierarchy.}
As an example of biological application, \cite{Ma2018using} demonstrate that neural model can simulate cellular growth almost as accurately as laboratory experiments.

\subsection{Protein-protein interaction prediction}\label{sec:protein-protein-interaction-prediction}

\begin{figure}[t]
\centering
\includegraphics[width=\linewidth]{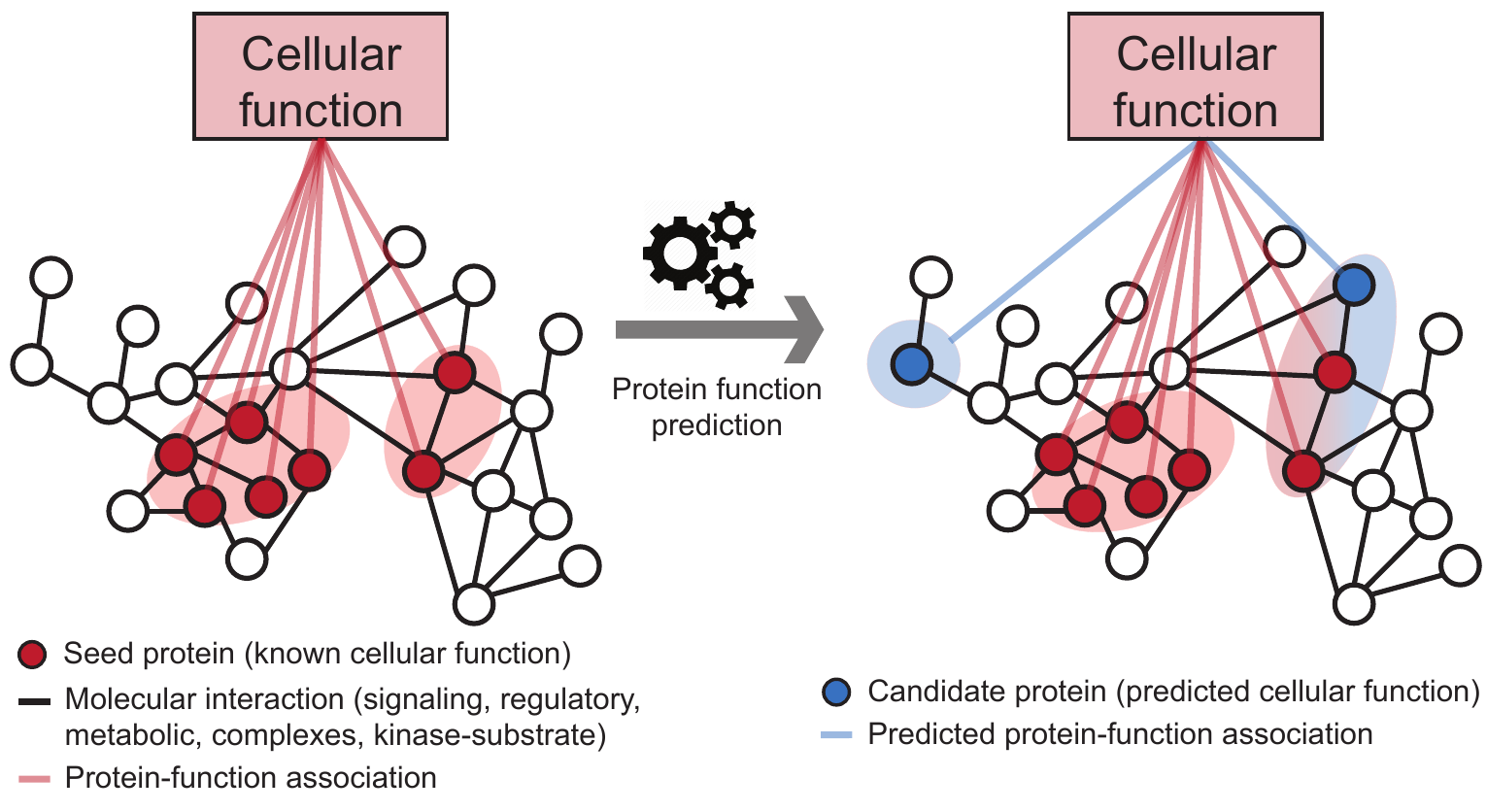}
\caption{\textbf{A network-based approach to cellular function prediction.} Biological networks are a powerful representation for the discovery of interactions and emergent properties in biological systems, ranging from cell type identification at a single-cell level to disease treatment at a patient level. Fundamental to biological networks is the principle that genes involved in the same cellular function or underlying the same phenotype tend to interact~\cite{Cowen2017network}. This principle has been used many times to combine and to amplify signals from individual genes, and has led to remarkable discoveries in biology. For example, network-based methods for protein function prediction~\cite{Li2010genome,Blatti2016characterizing,Zitnik2017ohmnet,Liu2017inferring} often use a heterogeneous protein-protein interaction network and conduct a large number of random walks on the network that are biased towards visiting known proteins associated with a specific function. These methods then calculate a score for each protein representing the probability that a protein is involved in a given cellular function based on how often the protein's node in the network is visited by random walkers.}
\label{fig:function-ppi-prediction}
\end{figure}

One major strategy to study cellular phenotype and function is to analyze networks of physical interactions between proteins.
These physical protein-protein interaction (PPI) networks carry out the core functions of \rev{cells since} interacting proteins tend to be linked to similar phenotypes and participating in similar functions~\cite{Menche2015}.
Protein-protein interactions also orchestrate complex biological processes including signaling and catalysis (\autoref{fig:function-ppi-prediction})~\cite{Cowen2017network}.

With the recent advances in experimental techniques, the number of identified PPIs keeps increasing~\cite{Rolland2014proteome}.
However, we are still far from \rev{the complete} knowledge of PPIs and their characterization at the network level.
Computational methods to predict PPIs have thus recently become popular due to the significant increase in other types of protein data, such as protein sequence and structural information, which is indicative of PPIs.

Proteins can interact with or co-localize with a variety of other biomolecules and can form stable complexes.
These complexes can bind to DNA, alter gene expression, and alter cell phenotype.
A predictive method by Jansen \etal~\cite{Jansen2003AData} improves analyses based on pull-down assays, which experimentally find proteins interacting with an input protein.
However, these assays tend to be noisy and are often incomplete.
To address this issue, Jansen \etal's~\cite{Jansen2003AData} method uses Bayesian inference across pairs of interacting proteins from a variety of datasets, along with transcriptomic and essentiality information to find complete interaction networks.
Another example is ChromNet~\cite{Lundberg2016ChromNet:Data}, which predicts PPIs among chro\-ma\-tin-inter\-acting proteins such as transcription factors using epigenomic data.
It does this by identifying conditional dependence structures between proteins present at specific genomic regions.
In another example~\cite{Drew2017integration}, over 9,000 mass spectrometry protein interaction datasets from a variety of human and animal cells and tissues were combined into a comprehensive map of human protein complexes and predict PPIs.
Interestingly, the combined map revealed thousands of PPIs that were not identified by any individual mass spectrometry experiment, thus demonstrating the value of data integration.
This analysis was accomplished by a network-based protein complex discovery pipeline.
The computational pipeline first generated an integrated protein interaction network using features from all input datasets.
To predict PPIs, the approach trained a protein interaction classifier based on support vector machines (SVMs). To predict protein complexes, the approach then employed a Markov clustering algorithm for graphs and optimized the clustering parameters relative to a training set of literature-curated protein complexes.

\section{Computational pharmacology}\label{sec:drug}

\begin{table*}[t]
  \centering
  \begin{tabular*}{\linewidth}{r@{\extracolsep{\fill}}p{\linewidth*5/7}}
    \toprule
    \textbf{Term} & \textbf{Description} \\
    \midrule
    drug discovery & The process through which potential new drugs are identified. It currently takes \rev{13--15} years and between \rev{\$2 billion} and \$3 billion on average to get a new drug on the market~\cite{Scannell2012diagnosing}. \\
    side effect & Secondary, typically undesirable effect of a drug, \rev{\eg adverse} drug reaction.\\
    medical indication & The use of a drug for treating a disease, \eg, insulin is indicated for the treatment of diabetes. \\
    drug-protein binding & The formation of a drug-protein complex. It describes the ability of a protein to form bonds with a drug in the human body. For example, if a drug is 95\% bound to a protein and 5\% free, \rev{then 5\% of the drug is active in the human body and causes pharmacological effects}.  \\
    \rev{target protein} & Protein that \rev{is addressed and controlled by a chemical compound}. Modulation of \rev{a target protein} can have a therapeutic effect. \rev{Target protein is {\em druggable} when it binds with} high affinity to a drug.\\
    drug-target interaction & A drug interacting with a target protein in the human body and affecting the protein's activity.\\
    drug-drug interaction & Phenomenon, in which the activity of one drug changes, favorably or unfavorably, if taken with another drug. It is often defined through the concepts of synergy and antagonism~\cite{Yeh2009drug}.\\
    protein-ligand binding & The process by which a ligand (usually a molecule) produces a signal by binding to a site on a target protein.\\
    drug combination & Combinatorial therapy that involves a concurrent use of multiple medications.\\
    structural interaction fingerprint & Binary vector representation of 3D structural information about protein-ligand binding. \\
    on-target side effect & Side effect that results from affecting the desired target protein of treatment. \\
    off-target side effect & Side effect that results from an unwanted interaction of a drug with other proteins.\\
    \bottomrule
  \end{tabular*}
  \caption{\textbf{Glossary for computational pharmacology.}
    Terms referenced in this section.}
  \label{tab:S9-glossary-terms}
\end{table*}

The goal of computational pharmacology is to use data to predict and better understand how drugs affect the human body, support decision making in the drug discovery process, improve clinical practice and avoid unwanted side effects (for an excellent review, see \cite{Li2015,Hodos2016}).
%
The properties of drugs and their interactions with the human body can be described in a variety of ways and measured at the physicochemical, pharmacological, and phenotypic levels.
One can measure the physicochemical properties of a drug, such as chemical structure, melting point, or hydrophobicity.
One can also measure interactions between a drug and its \rev{target proteins} by quantifying binding strength, kinetic activity, and the change in a cellular state or gene expression.
Furthermore, one can use phenotypic data, such as information about diseases that a particular drug treats, drug side effects, and interactions of a drug with other drugs.
Such data lend themselves to mathematical representations, which are then analyzed to guide drug discovery and {\em in vivo} experiments in a laboratory.

\subsection{Drug-target interaction prediction}\label{sec:drug-target-interaction-prediction}

At the most basic level, drugs have an impact on the human body by binding with \rev{target proteins} and affecting their downstream activity.
Identification of drug-target interactions is thus important for understanding key properties of drugs, including drug side effects, therapeutic mechanisms, and medical indications.
Traditional prediction of drug-target interactions uses {\em molecular docking}~\cite{Donald2011algorithms}, an approach that combines 3D modeling and computer simulation to dock a candidate drug into a protein-binding pocket and then score the likelihood of the pair's interaction.
This approach provides insights into the structural nature of the interaction, however, the performance of molecular docking is limited when the 3D structures of target proteins are not available.
As molecular docking can be computationally very demanding, {\em ligand-based methods}~\cite{Keiser2007relating} have emerged as an alternative approach to drug-target interaction prediction.
A ligand-based approach specifies an abstract model of chemical properties that are considered important for the interaction with the chosen target protein and then it aligns and scores candidate drugs against this model.
However, ligand-based approaches perform poorly when the chosen target protein has only a small number of known binding ligands and the quality of the abstract model is low.

\begin{figure}[t]
\centering
\includegraphics[width=\linewidth]{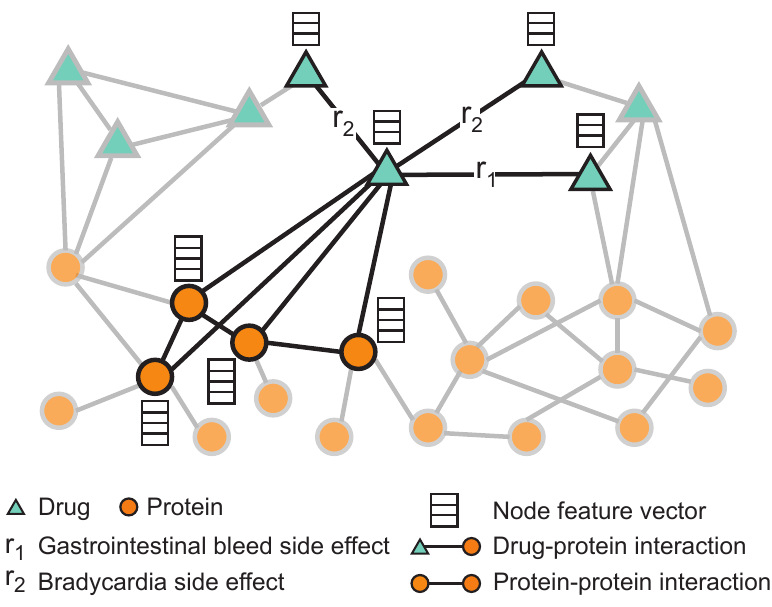}
\caption{\textbf{Drug-target and drug-drug interactions.} A heterogeneous network representation of drugs and proteins targeted by the drugs. In addition to interaction information, \eg, drug-drug interactions, drug-protein interactions, and protein-protein interactions (Section~\ref{sec:function}), each node in the network has a feature vector describing important biological characteristics of the node, \eg, drug's chemical structure, and protein's activity in tissues. Such networks are used to address two important tasks in computational pharmacology. The first is the prediction of drug-target interactions~\cite{Wang2014drug,Breinig2015chemical,Zong2017deep,Lee2017network}, which are fundamental to the way that drugs work and often provide an important foundation for other tasks in the computational pharmacology. The second is the prediction of drug-drug interactions~\cite{Vilar2014similarity,Cheng2014machine,Zitnik2016collective,Sridhar2016probabilistic}, which are fundamental to modeling drug combinations and identifying drug pairs whose combination gives an exaggerated response beyond the response expected under no interaction.  Zitnik \etal~\cite{Zitnik2018polypharmacy} use heterogeneous networks, such as the one shown in the figure, and develop a graph convolutional deep network approach to predict which side effects a patient might develop when taking multiple drugs at the same time.}
\label{fig:drug-drug-interactions}
\end{figure}

Many recent efforts focus on using machine learning for drug-target interaction prediction.
These efforts are based on the \rev{{\em guilt-by-association principle}}, a principle that similar drugs tend to share similar target proteins and vice versa.
Using this principle, prediction can be formulated as a binary classification task, which aims to predict whether a drug-target interaction is present or not.
This straightforward classification approach considers known drug-target interactions as positive labels and uses chemical structure of drugs and DNA sequence of \rev{target proteins} as input features (or kernels)~\cite{Bleakley2009supervised,Van2011gaussian,Wang2017networkassisted}.
Additionally, many methods integrate side information into the classification model, such as drug side effects~\cite{Campillos2008,Mizutani2012relating}, gene expression profiles~\cite{Iorio2010discovery}, drug-disease associations~\cite{Wang2014drug}, and genes' functional information~\cite{Yang2014drug}.
Such data provide a multi-view learning setting for drug-target interaction prediction~\cite{Gonen2013kernelized,Zhang2017drug}.
For example, \cite{Gonen2013kernelized} use kernelized matrix factorization and combine multiple types of data (\ie, views), each data type is treated as a different kernel, to obtain better prediction performance than single-kernel scenarios.
Another common approach is to represent multiple types of data as a heterogeneous network (\autoref{fig:drug-drug-interactions}) and predict \rev{target proteins} using random walks.
These methods use diffusion distributions to calculate a score for each node (protein) in the network, such that the score reflects the probability that the protein is targeted by a particular drug~\cite{Wang2014drug,Breinig2015chemical,Lee2017network}.
In addition to random walks, one can use meta-paths~\cite{Sun2011pathsim} to extract drug and protein feature vectors from a heterogeneous network and then \rev{feed} them into a classifier~\cite{Fu2016predicting}.

However, hand-engineered features, such as meta-paths, often require expert knowledge and intensive effort in feature engineering and can thus prevent methods from being scaled to large datasets.
For these reasons, matrix factorization algorithms are used to learn an optimal projection of a heterogeneous network into a latent feature space.
The learned latent space \rev{is used to infer a} drug-target network via a sequence of matrix operations and the resulting drug-target network is used to predict drug-target interactions~\cite{Zheng2013collaborative}.
A potential limitation of \rev{classic matrix} factorization is that \rev{it} takes as input a homogeneous network and thus one needs to collapse a heterogeneous network into a homogeneous one, discarding potentially useful information.
This limitation is overcome by multi-view, collective, and tensor factorization approaches to drug-target interaction prediction~\cite{Gonen2013kernelized,Narita2012tensor,Zitnik2016collective}.
In addition to \rev{using matrix factorizations, which are shallow feature learning algorithms,} one can use deep feature learning algorithms, such as deep autoencoders~\cite{Hinton2006reducing}\rev{.}
These algorithms generate a feature vector for every drug and protein in the dataset.
Using the learned drug and protein features, the method finds the best projection from the drug space onto the protein space such that the projected feature vectors of drugs are geometrically close to the feature vectors of proteins that are targeted by these drugs~\cite{Zong2017deep}.
The projection is learned to minimize prediction error on a training dataset of drug-target interactions~\cite{Luo2017network}.
After model training, the method predicts target proteins for a particular drug by ranking the proteins based on their geometric proximity to the drug's vector in the projected space.

\subsection{Drug-drug interaction and drug combination prediction}

The use of drug combinations is a common treatment practice.
Many patients take multiple drugs at the same time to treat complex diseases or co-existing conditions~\cite{Han2017synergistic}.
A drug combination consists of multiple drugs, each of which has generally been used as a single effective medication in a patient population~\cite{Jia2009mechanisms}.
Since drugs in a drug combination can modulate the activity of distinct proteins, drug combinations can improve the therapeutic efficacy by overcoming the redundancy in underlying biological processes~\cite{Sun2015combining}.
While the use of multiple drugs may be a good practice for the treatment of many diseases, a major consequence of a drug combination for a patient is a much higher risk of side effects which can be due to drug-drug interactions~\cite{Zitnik2014drug,Woo2017integrative}.
Such side effects can emerge because the activity of one drug may change if taken with another drug.
This means that a combination of drugs leads to an exaggerated response in patients that is over and beyond the response we would expect under no interaction.

Drug-drug interactions are one of the major concerns in drug discovery.
They are extremely difficult to identify manually because there are combinatorially many ways in which a given combination of drugs can clinically manifest and each combination is valid in only a certain subset of patients.
\rev{Furthermore, it is practically impossible} to test all possible pairs of drugs~\citep{Chen2016}, and observe side effects in relatively small clinical testing.
Given the large number of drugs, experimental screens of pairwise combinations of drugs pose a formidable challenge in terms of cost and time.
For example, given $n$ drugs, there are $n(n-1)/2$ pairwise drug combinations and many more higher-order combinations.
Furthermore, unwanted side effects are recognized as an increasingly serious problem in the health care system affecting nearly 15\% of the U.S. population~\cite{Kantor2015}.
To address this combinatorial explosion of candidate drug combinations, computational methods were developed to identify drug pairs that potentially interact~\citep{Ryall2015}.

Drug-drug interactions are defined through the concepts of synergy and antagonism~\citep{Loewe1953,Lewis2015} and are quantified biologically by measuring the dose-effect curves~\citep{Bansal2014,Takeda2017} or cell viability~\citep{Huang2014,Huang2014scirep,Sun2015,Zitnik2016,Chen2016,Chen2016synergy,Shi2017}.
Computational methods use these measurements to identify combinations of drugs, most often pairs of drugs, that potentially interact.
These methods predict drug-drug interactions by estimating the scores representing the overall strength of an interaction for a drug pair.
Existing approaches are classification- or similarity-based.
{\em Classification-based approaches} consider drug-drug interaction prediction as a binary classification problem~\citep{Cheng2014,Huang2014scirep,Zitnik2016,Chen2016,Shi2017,Zheng2017attention}.
They use known interacting drug pairs as positive examples and other drug pairs as negative examples.
The methods first obtain a feature representation of each drug pair.
For example, they use a linear or nonlinear dimensionality reduction algorithm on each data type to derive a feature vector for each drug~\cite{Zitnik2016,Zhao2016drug}, followed by an aggregation of feature vectors of individual drugs to obtain integrated feature vectors of drug pairs.
Finally, the methods train a binary classifier, such as logistic \rev{regression classifier}, support vector \rev{machine}, or neural network on feature representations of drug pairs.
In contrast, {\em similarity-based approaches} assume that similar drugs have similar interaction patterns~\citep{Gottlieb2012,Vilar2012,Huang2014,Li2015,Zitnik2015data,Sun2015,LI2017}.
These methods combine different kinds of drug-drug similarity measures defined on drug chemical substructures, structural interaction fingerprints, drug side effects, off-target side effects, and connectivity of molecular targets.
The methods aggregate similarity measures through clustering or label propagation \rev{to predict new} drug-drug interactions~\citep{Zhang2015label,Ferdousi2017,Zhang2017}.

Moving beyond predicting the chance of drug-drug interaction occurrence, recent methods identify how \rev{a given drug} pair manifests clinically within a patient population~\cite{Zitnik2018polypharmacy,Ma2018drug,Ryu2018deep}.
These methods use molecular, drug, and patient data to predict side effects associated with pairs of drugs.
For example, Decagon~\cite{Zitnik2018polypharmacy} constructs a multimodal graph of protein-protein interactions, drug-protein interactions, and drug-drug interactions (\autoref{fig:drug-drug-interactions}).
The approach represents each type of side \rev{effect} as a different edge type in the multimodal graph.
Decagon uses the graph to develop a graph convolutional neural network, a type of neural network designed for graph data~\cite{Hamilton2017review}, \rev{to predict side effects} of drug pairs.

\subsection{Drug repurposing}

\begin{figure}[t]
\centering
\includegraphics[width=0.9\linewidth]{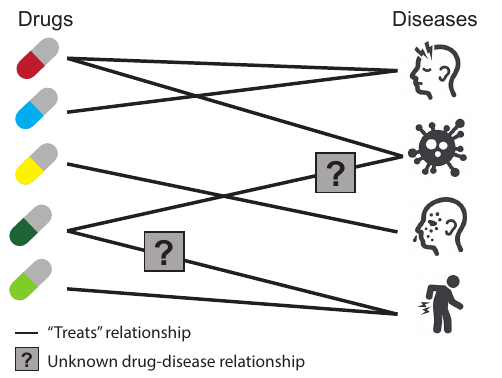}
\caption{\textbf{Drug repurposing.} \rev{Drug} repurposing uses computational methods to find new uses for existing \rev{drugs~\cite{Li2015,Hodos2016}.} Given a disease, the task is to predict drugs (\eg, among all drugs approved for use by the U.S. Food and Drug Administration) that might \rev{treat} that disease. Integrative methods for drug repurposing comprise similarity-based methods~\cite{Gottlieb2011predict}, network \rev{approaches}~\cite{Wang2014drug,Himmelstein2017systematic,Luo2017network}, and matrix factorization~\cite{Zhang2014towards}.}
\label{fig:drug-repurposing}
\end{figure}

Drug repurposing (also called ``drug repositioning'', \autoref{fig:drug-repurposing}) \rev{seeks} to find new uses for known drugs as well as for novel molecules.
Fundamental to drug repurposing are the following two observations.
First, many drugs have multiple \rev{target proteins}~\cite{Guney2016network} and hence a multi-target drug might be used for more than one purpose.
Second, different diseases share genetic factors, molecular pathways, and \rev{symptoms}~\cite{Zitnik2013discovering,Menche2015} and hence a drug acting on such overlapping factors might be beneficial to more than one disease.

At a high level, drug repurposing approaches can be categorized into four groups: (1) methods that predict new uses for existing drugs on the basis of protein target interaction networks~\cite{Li2009building,Wu2013network,Cheng2012prediction,Zhao2012co,Luo2017network}, (2) methods that make predictions by analyzing gene expression activation following various drug treatment regimes~\cite{Sirota2011discovery,Stanfield2017drug}, (3) methods that make predictions based on drug side effects~\cite{Fung2013extracting,Zhang2013exploring,Kuhn2013systematic,Wang2014exploring}, and (4) methods that consider a variety of disease similarity and drug similarity measures, each capturing a different type of biomedical knowledge~\rev{\cite{Wang2014drug,Gottlieb2011predict,Zhang2013computational,Li2013pathway,Yu2017prediction,Luo2016drug,Himmelstein2017systematic}.}

For example, \cite{Wang2013drug,Wang2014drug,Luo2016drug,Luo2017network} used random walks on a heterogeneous similarity network to rank candidate drugs for a given disease.
In another example, \cite{Luo2016drug} designed similarity measures to construct a drug-drug similarity network, a disease-disease similarity network and a drug-disease interaction network, and then used random walks to predict medical indications.
The method is based on the observation that similar drugs are used to treat similar diseases.
Along similar lines, the work of \cite{Gottlieb2011predict,Zhang2013computational} used multiple types of drug-drug and disease-disease similarity measures and combined them via a large-margin method or logistic regression to solve the drug repurposing task.

\section{Disease subtyping and biomarker discovery}\label{sec:patient}

\begin{figure}[t]
\centering
\includegraphics[width=\linewidth]{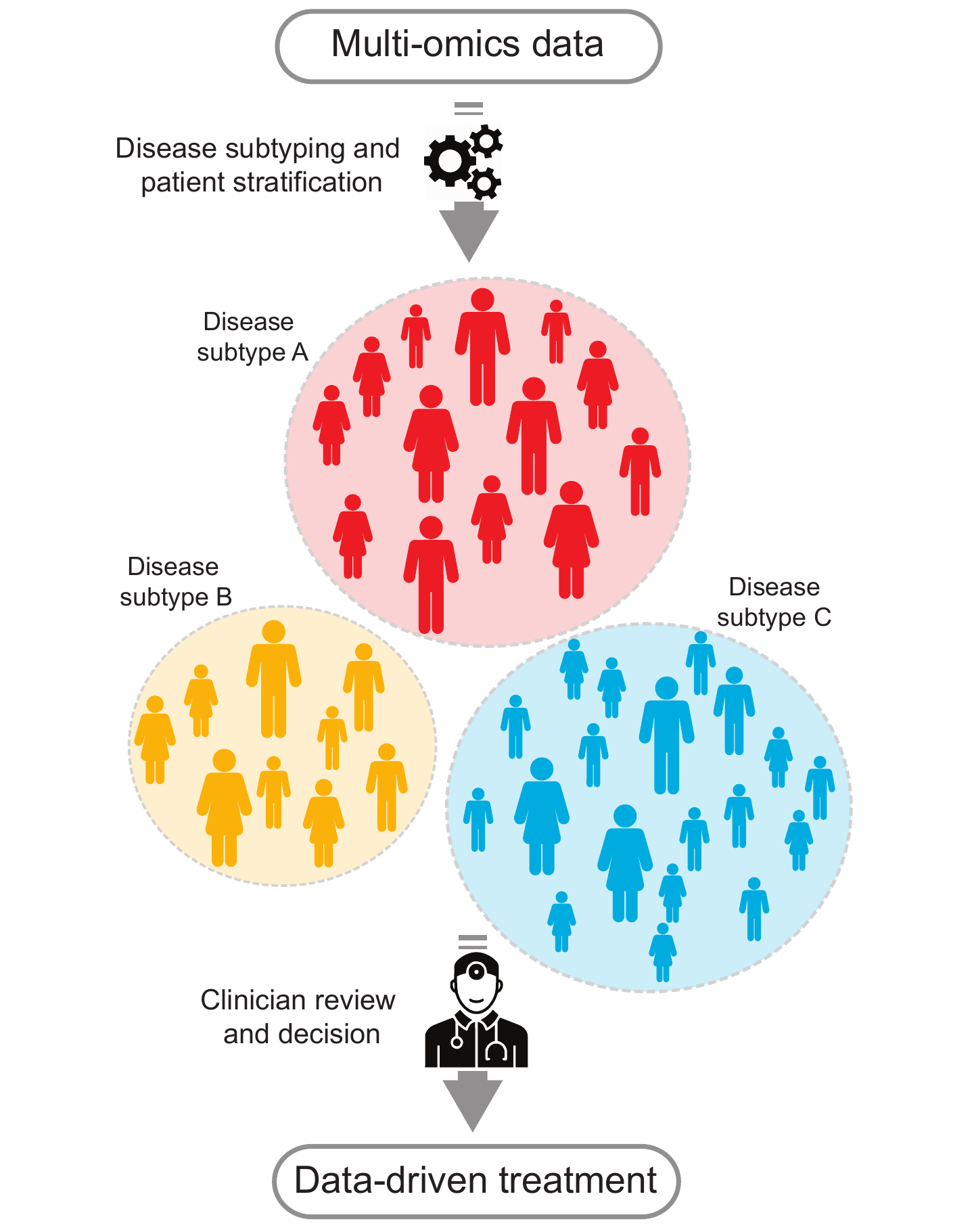}
\caption{\textbf{Disease subtyping.} Many diseases are heterogeneous. {\em Disease subtyping} stratifies a heterogeneous group of patients with a particular disease into homogeneous subgroups, \ie, subtypes, based on clinical, molecular, and other types of patient features. Accurate clustering of patients into subtypes is an important step towards personalized medicine and can inform clinical decision making and treatment matching.}
\label{fig:disease-subtyping}
\end{figure}

\rev{Many diseases} are characterized by incredible heterogeneity among patients. This includes many common diseases of which neuropsychiatric and autoimmune disorders (\eg, Au\-tism Spectrum Disorder (ASD), Attention Deficit Hyperactive Disorder (ADHD), Obsessive Compulsive Disorder (OCD), ar\-thritis, lupus, chronic fatigue syndrome \rev{(CFS))} are among the most diverse. This means that individuals present at the clinic with widely ranging symptoms. ASD patients, for example, range from those with mild behavioral challenges to inability to speak; arthritis can affect a very particular type of joint or present itself systemically, affecting multiple organs and tissues. For a lot of common diseases, there exist classifications into \emph{subtypes} that can be distinguished clinically (\autoref{fig:disease-subtyping}).  Consequently, treatment maybe guided by that clinical distinction. On the other hand, diseases such as cancer present themselves, for example, as a solid mass in a given organ (\eg, lung, breast, stomach, etc) and clinically seem similar, however biopsy and the consequent cellular profiling revealed that these masses may widely differ, conferring different risks and prognoses for the patients. A good example is breast cancer, where at least four different subtypes are currently distinguished in the clinic based on gene expression biomarkers (Luminal A and B, Her2+, Triple Negative/Basal-like). Further research on breast cancer has sho\-wn that there maybe closer to ten subtypes \cite{Curtis:2012} or even more. It thus appears that there is both clinical and biological heterogeneity across multiple diseases. The cancer scenario tells us that clinical and biological subcategorizations of disease might not agree, indeed, the symptoms with which breast cancer patients present in the clinic are not indicative of their molecular subtypes.

Determining subtypes computationally presents a challenge. In theory, subtyping a disease means identification of homogeneous subgroups of patients, \ie, clustering, yet we see that in practice, clustering of different types of patient information (clinical vs molecular data) leads to different subgroupings of patients. This inconsistency is not only present between molecular and clinical data, it is also present among molecular subtypes. For example, \cite{Cavalli:2017} showed that clustering of gene expression vs methylation of medulloblastoma (brain cancer) patients resulted in inconsistent subgroups which were resolved using \emph{integration} of gene expression and methylation. Another example, is the case of glioblastoma \rev{multiforme} (GBM), a very aggressive adult-onset brain cancer. An earlier analysis combining gene expression and \rev{copy number variations (CNVs)} yielded two subtypes \cite{Nigro:2005}, whereas a later analysis, driven primarily by gene expression analysis, yielded 4 subtypes \cite{Verhaak:2010}. Interestingly, that while methylation data was available in \cite{Verhaak:2010}, it was used only to explain the clusters obtained with gene expression and thus it was found to be uninformative. Analysis that used methylation as the driving signal identified a very prominent and now \rev{well-recognized} \textit{IDH1} subtype, a mutation that leads to \rev{hypermethylation across} the genome that corresponds to a younger subpopulation of GBM patients with better clinical prognosis. To summarize, analyzing each of the molecular data types independently resulted in inconsistent findings that were difficult to consolidate. These examples illustrate the importance of data integration to identifying subtypes. Indeed, the more completely we can define the patients, the more faithful and hopefully, clinically relevant, will our subtypes be.

Many methods for data integration have been developed with the purpose of identifying disease subtypes. The simplest commonly used method is the concatenation of all the available data types and then clustering patients using the long concatenated vectors. The problem with this approach is that it completely disregards the structure present in each of the datasets, thus diluting the often weak signal even further. Another simple method that avoids this issue is Cluster-Of-Cluster-Assignments (COCA), which was originally developed to define subclasses in \rev{The Cancer Genome Atlas (TCGA)} breast cancer \rev{patient cohort}~\cite{Koboldt:2012}. COCA first clusters patients according to each of the individual data types and then takes these assignments as binary vector inputs and re-clusters patients according to those vectors thus providing consensus. The problem with this assignment is that it is mostly driven by the common signal across all data types, not making use of the complementary information potentially provided by the different data types. This approach was used by the TCGA to integrate five data types including mRNA, DNA methylation, \rev{reverse phase protein array} (RPPA), CNV and miRNA data across 12 cancer types and they successfully re-identified majority of the cancer types \cite{Hoadley:2014}. The reality, however, is that one can obtain very similar accuracy by clustering these samples using mRNA only. The problem was with the borderline cases that multiple types of data disagreed on. COCA was unfortunately not particularly useful for most of those cases.

There are many more sophisticated approaches that try to capture internal structure, latent dimensions and non-linearity. For example, iCluster is a Gaussian latent variable model with sparsity regularization in \rev{a Lasso-type} optimization framework \cite{Shen:2009}. The main assumption behind this method is that there exists \rev{a latent} space that captures the true subgrouping of the patients. Each of the different data types are then used jointly to estimate this latent space. This method was applied to identify 10 breast cancer subtypes from the METABRIC cohort \cite{Curtis:2012}. In our experience, iCluster results tend to be dominated by the strongest single data type signal. Another drawback of iCluster is that it cannot naturally handle thousands of variables (genes)\rev{; thus} gene pre-selection has to be applied to the data first. This pre-selection imposes a bias and if the pre-selected features do not contain signal relevant to the true subgroups, it will be hard to impossible to recover them in the post-selection integration. Patient Specific Data Fusion (PSDF) \cite{Yuan2011patient} is another latent variable approach. PSDF is a nonparametric Bayesian model for discovering subtypes by combining gene expression and copy number variation. PSDF estimates a latent variable per patient, minimizing samples on which the combined data types contradict each other. While a powerful non-parametric framework, PSDF suffers from high computational costs due to the necessity to infer a large number of parameters and the restriction to combine only two data types.

Another type of methods for integrating data to identify subtypes is network-based. An example of such an approach is Similarity Network Fusion (SNF) \cite{Wang2014similarity}. Instead of trying to combine data in the original measurement space that are hard to calibrate and compare across a variety of data types, SNF combines data in the patient similarity space. In short, SNF consists of two \rev{steps;} first it creates a similarity patient network for each of the available data types and once all the networks are constructed, it combines these networks in an iterative non-linear fashion relying on an idea of extension of random walks across multiple graphs. SNF was shown to outperform the \rev{above methods} \cite{Wang2014similarity} on five cancers and has subsequently been applied outside of cancer to combine images and clinical data as well as a variety of lab tests across multiple diseases \cite{Cavalli:2017,Vega:2018,Zizzo:2018,Stefanik:2017,TCGA:2017}. \rev{In spirit, SNF is similar to Multiple} Kernel Learning (MKL), which can also be used to construct and combine similarities \cite{Huang:2012}. The main difference between SNF and MKL is the linear nature of MKL which hurts its performance during integration as shown in \cite{Wang2014similarity}. While there are not as yet many methods that perform subtyping using network fusion, a short review on the topic can be found in \cite{Pai:2018}.

When it comes to biomarker discovery, there is a myriad of papers, however, when it comes to the integrative analysis, the number of approaches identifying truly integrative biological markers is sparse. One of the early and very interesting approaches is called PAthway Recognition Algorithm using Data Integration on Genomic Models (PARADIGM) \cite{Vaske:2010}. In a nutshell, this method models activity levels of each gene, which are represented as latent variables. The method relies on a large public network of genes, including activating and inhibitory interactions. This network is then transformed into a Bayes \rev{network,} for which the following biological assumptions are followed: for each gene, \rev{copy number alteration (CNA)} affects expression, which affects protein levels, which affect the latent protein activity. This graph represents the reference (normal) state. Given the data for a particular disease, a joint posterior distribution is computed for all latent activity nodes. By comparing pre- and post-activity levels, PARADIGM obtains a quantitative measure of the alteration induced by the disease. This approach was applied in the pancancer study \cite{Hoadley:2014} and biologically relevant dis-regulations were identified.

\section{Challenges and future directions}\label{sec:discussion}

There are great opportunities at the intersection of machine learning and biomedical data integration. However, there are equally great challenges that need to be overcome. In particular, the days of studying biomedical datasets in isolation and independently of each other are slowly coming to an end and the reductionist paradigms of looking for `low-hanging fruit' (\ie, \rev{a single variable that would fully explain a trait}) are becoming less prevalent. The realization that performing all analyses within only one data type \rev{can limit the potential for discovery} of new biomedical insights has led to the development of many new ideas and methods for integrating biomedical data. However, these approaches are only in their beginning and little is known about key principles of their optimal design. In addition, gold standard methods for many biomedical \rev{questions}, such as identifying noncoding DNA variants (Section~\ref{sec:noncoding}), multi-omics profiling of \rev{cells} (Section~\ref{sec:singlecell}), and stratifying patient populations (Section~\ref{sec:patient})\rev{,} are only emerging. Furthermore, \rev{combinations of heterogeneous data and new machine learning methods has enabled us} to ask fundamentally new biomedical questions.

There are many directions available to take on these challenges. Below, we highlight outstanding problems and opportunities that need to be addressed to fully realize the potential of machine learning for integrating biomedical data.

\subsection{Combining mixed-technology data}

The structure and \rev{distribution} of data generated by different technologies (\eg, gene expression data generated by a sequen\-cing-based \rev{versus} an array-based technology~\cite{Wang2014concordance}) can be very different and it is challenging to combine such data. Data normalization is thus an essential first step when analyzing mixed-technology data. Furthermore, there is a deluge of \rev{different assays} (\eg, Table~\ref{tab:epig-glossary-assays} and Section~\ref{sec:singlecell}) and \rev{appropriate normalization of data} derived from \rev{these assays} prior to downstream \rev{analysis} remains a major challenge. Normalization is \rev{important because} it can adjust for unwanted biological and technical noise that can mask the signal of interest. For example, one widely used normalization strategy in single-cell transcriptomics is glo\-bal scaling~\cite{Vallejos2017normalizing} that removes cell-specific biases by scaling gene expression measurements within each cell by a constant factor. There are many opportunities for moving data normalization approaches forward by using next-generation machine learning methods. For example, one could use generative adversarial networks (GANs) to generate data with the properties of real data and then use the created data to normalize the real data. Future approaches may include {\em integrated strategies}, where normalization is intrinsic to a specific type of analysis (\eg, \cite{Hiranuma2018aicontrol}), and {\em generic tools}, which normalize the data that can then be used as input to any downstream analysis (\eg, \cite{Bacher2017scnorm,Taroni2017cross,Wang2018network}).

\subsection{Multi-scale and higher-order approaches}

A central goal of computational biology is to assemble a predictive model of a cell that would be able to predict a range \rev{of phenotypes} and answer biological questions.
To be able to predict \rev{many phenotypes}, rather than only one type of outcome, we need to understand how phenotypes are interrelated with each other.
Here, multi-scale models come into play because the cell is organized in a hierarchical manner, both in 3D structure and in function~\cite{Carvunis2014siri}.
Similarly, \rev{higher-order} structure and function of the cell might emerge from many molecular measurements and interaction datasets if only one could figure out how to combine these measurements properly.
A multi-scale predictive model \rev{of the cell is} a very general framework, but whether it can capture the full extent of biological complexity remains to be seen.
Furthermore, it is not clear how to combine or extrapolate cell models to the scale of an organism (\ie, human patient).
This gap between \rev{cell models} and \rev{organismal models} poses fundamentally new challenges that must \rev{eventually be met}.
Moreover, because \rev{parameters of most current machine learning models are fixed} after the model is trained, such \rev{models are} incompatible with biological evolution.
First critical steps to address these challenges have already been taken.
For example, recent advances in the theory of multi-level graphs and network motifs enabled us to study, for example, higher-order organization of gene regulatory networks~\cite{Milenkovic2008uncovering,Benson2016higher} and multi-layer nature of ecological systems~\cite{Pilosof2017multilayer}.
Furthermore, these challenges present an excellent opportunity for next-generation machine learning algorithms, such as those based on deep representation learning and topological data analysis, to develop multi-scale~\cite{Zitnik2017ohmnet} and higher-order~\cite{Rizvi2017single} models of a cell, and eventually of a human patient.

\subsection{Interpretability and explainability}

The black-box nature of many  machine learning methods presents an additional challenge for biomedical applications.
It \rev{is  difficult to interpret the output of such methods} from a biomedical perspective, a challenge that limits \rev{methods' utility} for providing insights\rev{.}
This is especially the case for advanced \rev{methods} such as deep neural networks that transform the input data in such a way that it can be difficult to determine the relative importance of each feature or whether a feature is positively or negatively correlated with the outcome.
Understanding black-box predictions is an open challenge in machine learning, with great attention being given to the interpretation of how a particular model relates the input to its output~\cite{Ribeiro2016should,Lundberg2017unified,Arpit2017closer,Koh2017understanding}.
There is a critical need to develop means to \rev{transform {\em black-box methods} into {\em white-box methods}} that can be opened up and interpreted meaningfully\rev{.}
An early application of explainability in biomedicine includes \cite{Lundberg2017explainable}, an approach that integrates high fidelity data from a hospital's information management system (\eg, data from patient monitors and anesthesia machines, medications, laboratory results, and electronic medical records) to predict the risk of hypoxemia during surgery and explain the patient- and surgery-specific factors that lead to that risk.
In a similar way, \cite{Ma2018using} used a neural network and integrated \rev{prior biological knowledge from \rev{the GO}~\cite{Ashburner2000gene} into the neural model}.
A particular genotype-phenotype association could then be explained by a hierarchy of \rev{cellular systems, which was identified as a neural activation map.}

\subsection{Integration of self-reported, lifestyle, and ecological data}

While the cost and speed of generating genomic data have come down dramatically in recent years, advances in the collection of phenome data (\ie, the set of all phenotypic information for a single organism or individual, see Section~\ref{sec:patient}) have not kept pace.
To begin to address the phenomics challenge, new research \rev{is needed to facilitate both broad} and deep phenotyping and maximize the utility of gathered data while minimizing the burden on individuals.
Although studies have traditionally used medical records \rev{as gold} standard information about medical conditions, emerging research \rev{considers} internet and mobile technologies as a viable method for broad phenotyping \rev{of} large populations.

Relative to medical record review, {\em internet-based phenotyping} can be \rev{fast. For example, Tung et al.~\cite{Tung2011efficient} assessed more than 20,000 people for 50 phenotypes, such as Crohn's disease, inflammatory bowel disease and diabetes, in approximately 12 months using only a small team of people.}
Emerging research has demonstrated the value of combining these self-reported data with genomic information about individuals.
For example, Hu et al.~\cite{Hu2016gwas} conducted a genome-wide association analysis of self-reported morningness (\ie, a morning person prefers to rise and rest early) and then analyzed the newly identified genetic variants using biological pathways.
Along similar lines, Hyde et al.~\cite{Hyde2016identification} recently used self-reported data from more than 300,000 individuals and combined them with a genome-wide association study to identify genetic variants associated with depression.
Furthermore, integration of other types of lifestyle and ecological data together with molecular information has a large potential to reveal new biological mechanisms.
For example, \cite{Smits2017seasonal} is an early work in this area that combined human gut microbiome data with lifestyle information.
The combined data revealed striking differences in gut microbial communities between seasons that depended on seasonal availability of different types of food.

\section{Conclusions}

Machine learning is becoming integral to modern biomedical research.
Importantly, \rev{approaches have emerged that can integrate data from many different biomedical datasets.}
These approaches aim to bridge the gap between our ability to generate vast amounts of data and our understanding of biomedical systems and thus reflect the intricate complexity of biology.
Ongoing methodological developments and emerging applications of machine learning promise an exciting future for biomedical data integration, although it is likely that no single method will perform best for all \rev{problems}.
Approaches thus need to be selected according to different types of domain-specific models, specific types of data, and different types of biomedical outcomes.
In this Review, we described various \rev{approaches that} can currently be implemented to perform powerful integrative analyses.
As integrative approaches \rev{become} more readily available, systems biology and systems medicine are likely to become a central computational strategy to generate new knowledge in biology and medicine.

\section*{Acknowledgments}

\noindent M.Z.~and J.L.~were supported in part by NSF IIS-1149837, NIH BD2K U54EB020405, DARPA SIMPLEX, Stanford Data Science Initiative, and \rev{the Chan} Zuckerberg Biohub. A.G.~was supported by the Natural Sciences and Engineering Research Council of Canada (Discovery Grant to A.G.) and the Canadian Cancer Society (Innovation Grant to A.G.). F.N.~and M.M.H.~we\-re supported by the Natural Sciences and Engineering Research Council of Canada (RGPIN-2015-03948 to M.M.H.). 

\bibliography{paper-fusion}

\end{document}